\documentclass[fleqn,10pt]{wlscirep}
\usepackage[utf8]{inputenc}
\usepackage[T1]{fontenc}
\usepackage{multirow}
\usepackage{amsmath}

\mathcode`*=\string"8000
\begingroup
\catcode`*=\active
\xdef*{\noexpand\textup{\string*}}
\endgroup
\title{Causal spillover effects of electric vehicle charging station placement on local businesses: a staggered adoption study}

\author[1,2,3]{M. Mavin De Silva}
\author[1,4]{Callie Clark}
\author[2]{Tadachika Nakayama}
\author[1,5*]{Takahiro Yabe}
\affil[1]{Center for Urban Science and Progress (CUSP), Tandon School of Engineering, New York University, Brooklyn, NY 11201, USA}
\affil[2]{Extreme Energy-Density Research Institute, Nagaoka University of Technology, Nagaoka, Niigata 940-2188, Japan}
\affil[3]{Department of Transport Management \& Logistics Engineering, Faculty of Engineering, University of Moratuwa, Katubedda 10400, Sri Lanka}
\affil[4]{Marron Institute of Urban Management, New York University, Brooklyn, NY 11201, USA}
\affil[5]{Department of Technology Management and Innovation, Tandon School of Engineering, New York University, Brooklyn, NY 11201, USA}

\affil[*]{takahiroyabe@nyu.edu}

\begin{abstract}
 Understanding the economic impacts of the placement of electric vehicle charging stations (EVCSs) is crucial for planning infrastructure systems that benefit the broader community. Theoretical models have been used to predict human behavior during charging events, however, these models have often neglected the complexity of trip patterns, and have underestimated the real-world impacts of such infrastructure on the local economy. In this paper, we design a quasi-experiment using mobile phone GPS location and EVCS deployment history data to analyze the causal impact of EVCS placement on visitation patterns to businesses. More specifically, we leverage the staggered placement of EVCSs in New York City and California Bay Area to match treated and control businesses that share similar characteristics including the business sector, location, and pre-treatment visitation count. By comparing three alternative matching strategies, we show that staggered adoption avoids selecting controls from non-treated clusters, and yields greater spatial overlap in dense urban areas. We find that EVCS installations significantly increase customer traffic, with effects concentrated in recreational venues in New York City and routine destinations such as groceries, pharmacies, and cafés in California Bay Area. Our results suggest that the economic spillovers of EVCSs vary across urban contexts and highlight the effectiveness of leveraging the staggered nature of adoption timings for evaluating infrastructure impacts in heterogeneous urban environments.
 \end{abstract}
\begin{document}

\flushbottom
\maketitle
\thispagestyle{empty}

% \noindent Please note: Abbreviations should be introduced at the first mention in the main text – no abbreviations lists. Suggested structure of main text (not enforced) is provided below.

\section*{Introduction}

Electric vehicles (EVs) are increasingly seen as a transformative solution in the quest for decarbonizing transportation and reducing global greenhouse gas emissions. Light-duty vehicles contribute approximately 80 percent of New York City’s total transportation emissions\cite{NYCDOT2021}, underscoring the need for cleaner alternatives. In recent years, light-duty EV prices have dropped to as low as \$18,875 when federal tax credits and state rebates are applied\cite{Kane2021}, and driving ranges have extended up to 400 miles per charge\cite{EVDatabase2022}. While the economic and performance aspects of EVs are improving, the accessibility and availability of electric vehicle charging stations (EVCSs) represent the next crucial challenge for widespread adoption of EVs. To alleviate these barriers and stimulate investments in the adoption of EVs, the US government has proposed a \$7.5 billion initiative to install 500,000 EV charging stations nationwide by 2030\cite{Wayland2021,Bellon2021}. With such a large and widespread investment and the early stage of the deployment of EV infrastructure, strategic site selection is important. The optimal strategy to place the chargers is an open question as there are many aspects of access and benefits to consider.

Studies have shown that strategically allocating subsidies and investments in EV charging infrastructure plays a critical role in promoting EV adoption\cite{Yu2022,Zhang2021}. However, such research often overlooks a key stakeholder: nearby businesses and amenities. Many EVCSs are located near retail destinations, providing access to other activities, such as shopping or dining, during charging times. Understanding how EVCSs create demand spillovers to surrounding businesses is essential to enable policymakers, EVCS providers, and business owners to guide policies and investments that create positive economic impacts in local communities. 

Previous studies have explored the positive effects of installing EVCSs on customer counts and spending in nearby businesses, highlighting their role as economic stimulators. To quantify the spillover effects of EVCS installation on consumer counts and spending, Zheng et al. (2024) match businesses located near EVCSs with comparable businesses located outside the treatment area of any EVCS, using propensity scores to control for factors that confound the placement of EVCSs and visitation patterns \cite{Zheng2024}. Previous work by Babar and Burtch (2024) argues that the distribution of EVCS is closely tied to interdependent socio-demographic and land-use characteristics of local populations, and therefore is not random \cite{Khan2022,Babar2024}. To address this strategic placement of EVCSs, their identification strategy leverages the staggered timing of treatment across physical retail establishments, where treatment is defined as the construction of a new Tesla Supercharger station in close proximity. The control group is drawn from never-treated establishments that operate under the same brand within the same county \cite{Babar2024}. However, as the density of deployed EVCS continues to increase, selecting controls from never-treated establishments can lead to limited spatial overlap between treated and control groups. In practice, treated businesses are often located near central business districts, while potential control businesses tend to be situated in lower-density peripheral areas. This spatial imbalance can introduce bias into the estimated treatment effects by violating the comparability assumptions required for credible causal inference. To address the limitations identified in prior studies \cite{Zheng2024,Babar2024}, we propose a novel approach for estimating the spillover effects of EVCSs on businesses that leverages the staggered adoption of EVCSs. This approach enables us to utilize businesses near EVCSs as control group units by analyzing the visitation patterns before the EVCS placement, leading to a larger number of available control units especially in high density areas and robust causal estimates.

Understanding how providers may strategically place EVCSs to maximize the benefits of deployment is key to unlocking the full potential of EV infrastructure and fostering vibrant and sustainable communities \cite{Babar2024,PYMNTS2022}. This study contributes to bridging this knowledge gap by leveraging the staggered installation of public EVCSs across two urban settings: New York City (NYC), characterized by high EVCS density, and California Bay Area, which has moderate urban density and high EV adoption rates. Our analysis using a staggered difference-in-differences (DID) approach shows that EV charging stations generate significant demand spillovers for nearby businesses, with a new EVCS generating an average increase of $\sim 28$ daily visits in NYC and $\sim 7$ in California Bay Area. Moreover, the spillover effects of EVCSs vary substantially across both regions: they concentrate on recreational or exploratory destinations in NYC, but in California Bay Area they are strongest for routine destinations such as grocery stores, pharmacies, and cafés. These differences reflect how local urban structure shapes EV charging related mobility. These results suggest that contingent factors, like the EVCS types, type of business and how close they are to each other, can influence how much businesses benefit from being near EVCS.

\section*{Results}

A key challenge in evaluating the causal impact of an EVCS installation is the potential endogeneity in their placement. EVCS providers often strategically deploy charging stations in locations where the benefits of EVCS are likely to be maximized\cite{Zheng2024,Babar2024}. This strategic placement could bias the results, as points of interest (POIs) near EVCS installations may inherently differ from those without nearby EVCSs. To address this concern, we employ a staggered adoption approach, which allows us to account for temporal and spatial variations in the timing of EVCS installations. Figure \ref{Figure 1}a illustrates the selection criteria for control and treatment POIs, focusing on a structured comparison framework. POIs within a 400 m radius of EVCS installations are designated as treated and control groups, with a minimum of 3 months between successive EVCS installations considered to ensure proper temporal alignment in treatment. In the matching process, we select POIs that are located in the same borough and fall under the same business category. Additionally, the control POI must be associated with a different EVCS than the one linked to the treated POI, and the EVCS serving the control POI must have been installed at least three months after the installation date of the treated POI's station. Among the set of candidate control POIs, we identify the control POI by selecting the POI with the smallest `pre-treatment difference' in visits. More specifically, we define the "pre-intervention difference" as the difference in average visits between 3-6 months before the intervention and the average visits between 1-3 months before the intervention. This temporal separation is critical, as it reduces the risk of confounding trends and ensures the independence of treatment effects across time. By leveraging this gap, the staggered DID approach isolates the causal impact of EVCS openings on nearby businesses, as shown in Figure \ref{Figure 1}a. The green line denotes POI \textit{j}, acting as a control for treated POI \textit{i}, which was subsequently impacted by a EVCS in a later date. The summary statistics for the variables about treatment, and control POIs are presented in Supplementary Tables S4 and S5. 

Several studies have used propensity score matching (PSM) to pair treatment POIs with control POIs, where control groups are selected from POIs without proximate EVCS installations during the study period\cite{Zheng2024,Babar2024,Ito2024}. However, in EVCS dense areas such as Manhattan and Brooklyn, the abundance of POIs near EVCS installations complicates the selection of an appropriate control group, posing challenges for conducting comparable studies\cite{Esmaili2024}. Although prior research has noted the strategic placement of EVCS to optimize their effectiveness, few studies have rigorously accounted for the inherent locational characteristics that influence EVCS deployment within causal inference frameworks\cite{He2016,Ren2019,Liang2023}. To address potential endogeneity, we compared the staggered adoption with the non-staggered matching approach \cite{Babar2024} proposed by Babar and Burtch (2024) and the non-staggered PSM approach \cite{Zheng2024} discussed in Zheng et al. (2024) across two urban settings: NYC, characterized by high EVCS density, and California Bay Area, which has moderate urban density and high EV adoption rates. This comparison evaluates whether ignoring the intrinsic characteristics of EVCS site selection introduces bias in causal inference. In the two non-staggered matching approaches, untreated POIs were designated as those located outside the 400m radius of any EVCS opened during the same period. The non-staggered matching approach was implemented using a two-step procedure, adapted from the staggered framework, to pair treated POIs with control POIs that exhibit similar characteristics. First, we performed an exact match based on the same borough \textit{b} and POI category \textit{c}, ensuring that each treated POI is paired with a control POI, thereby enhancing comparability between groups. Subsequently, we applied a nearest-neighbor algorithm to minimize the absolute pre-treatment difference, to ensure the parallel trend assumption. The summary statistics for the variables about treatment, and control POIs obtained under non-staggered matching approach are presented in Supplementary Tables S6 and S7. In the non-staggered PSM approach, we carefully select control POIs from the untreated POIs using PSM with a two-step approach. First, we conducted an exact matching based on the same borough \textit{b} and POI category \textit{c}, thereby enhancing comparability. Subsequently, we employ nearest-neighbor PSM using propensity scores estimated from a comprehensive set of covariates, following the approach proposed by Zheng et al. (2024) \cite{Zheng2024}. These characteristics include: (1) Built environment factors (population density, auto-oriented road/total road network density, and walkability) reflect accessibility and urban activity, as EVCS are often installed in densely populated, vehicle-accessible areas; (2) Socio-demographic variables (income, unemployment, and race composition) account for economic conditions and EV adoption potential; (3) POI-level variables, such as pre-installation customer counts, explains areas with high consumer activity, which are attractive for EVCS installation. The summary statistics for the variables about treatment, and control POIs obtained under non-staggered PSM approach are presented in Supplementary Tables S8 and S9. To explore the relationships between these covariates and treatment, we employ logistic regression, regressing the treatment variable on these covariates. For an effective evaluation of the staggered framework relative to non-staggered approaches, we focus on Manhattan, NYC, and San Francisco, which represent two distinct urban settings. We assess the covariate balance between treatment and control groups using graphical comparisons, as shown in Figure \ref{Figure 1}b. In the staggered adoption approach, covariate differences between treatment and control groups converge toward zero ($-0.1 < \text{covariate difference} < 0.1$) in both cities, demonstrating the robustness of the matching strategy. In contrast, the non-staggered matching approach, although providing a valid comparison at the metropolitan scale, exhibits weaker covariate balance in certain variables (outlined in black), with imbalances becoming more evident within the downtown areas. Figures S9 and S10 illustrate the covariate balance and distance between matched pairs in NYC and California Bay Area, respectively. 

The matched pairs developed in Figure \ref{Figure 1}c (Manhattan, NYC) and Figure \ref{Figure 1}d (San Francisco) through the staggered DID approach yielded a larger number of treated-control matches compared to both non-staggered approaches, indicating that incorporating treatment timing into the matching process improves spatial comparability. By calculating Moran’s I statistics (Figure \ref{Figure 1}c and \ref{Figure 1}d), we find that treated and control POIs are more spatially integrated under the staggered approach in both Manhattan, NYC and San Francisco, with lower Moran’s I values ($\sim 0.18$), indicating that comparisons are less influenced by geographic clustering. In contrast, the non-staggered approach exhibits much higher Moran’s I values ($\sim 0.71$), reflecting a higher risk of spatial bias as the treated and control groups have less spatial overlap\cite{Liang2023}. This pattern suggests that the spatial distribution of treated and control POIs exhibits significant spatial autocorrelation, underscoring the importance of accounting for spatial dependence when evaluating causal effects in urban settings. Following the matching process, we evaluated the distances between matched pairs using graphical representations. Supplementary Figure S9 (NYC) and Figure S10 (California Bay Area) show that the staggered approach achieves shorter matched-pair distances compared with the non-staggered approaches in high EVCS density areas. The mean differences in matched pair distances across methods are statistically significant ($p < 0.001$), indicating that incorporating later-treated POIs as controls enhances spatial overlap between treated and control groups.

\subsection*{Placement of public EVCS causes increase of visits to businesses}

We begin our analysis by evaluating the causal impact of public EVCS placement on nearby business activity using several estimation strategies. To examine the temporal evolution of these effects, we first implement an event study design (described in the Methods section). Figure \ref{Figure 2}a (NYC) and \ref{Figure 2}b (California Bay Area) show the estimated effects of an additional nearby EVCS on changes in customer visits. The results indicate that there are no statistically significant differences before installation, which supports the parallel trend assumption of the staggered DID framework. After installation, the effect becomes significantly positive, corresponding to an average increase of 31.05 customers $(\text{p-value} < 0.001)$ in NYC and 7.22 customers in San Francisco $(\text{p-value} < 0.001)$. These findings suggest that EVCS installations contribute to measurable increases in local business activity. It is noteworthy that in NYC, the marginal effect declines more rapidly in the later stage of the study period than in California Bay Area, likely reflecting faster saturation in denser commercial areas where EVCS utilization per port decreases as installations expand. In contrast, California Bay Area’s more dispersed urban layout allows the positive effects to persist longer as travel and charging behaviors adjust more gradually. Further research is necessary to pinpoint the precise reasons. Given the much higher density of EVCS installations in NYC, we focus our extended robustness checks on this urban context. The staggered DID approach estimates that the average treatment effect of an EVCS placement on a business is 28.52 customers $(\text{p-value} < 0.001)$, as shown in the first column of Table \ref{tab:my_table}. To ensure that this estimate is not affected by our choice of model parameters, we conducted a series of robustness checks, shown in Table \ref{tab:my_table}. 

Several studies have argued that EVCSs are systematically installed in locations that are more likely to benefit co-located retail establishments \cite{Zheng2024,Babar2024}. However, if businesses in our sample were actively influencing the arrival of EVCSs, it is reasonable to assume that the establishments located closest to the EVCS would be the most likely to engage in such behavior. We find that, by dropping those establishments that are located closest to the treated EVCS, our DID estimations remain unchanged $(\beta = 24.50,\ \text{p-value} < 0.001)$, indicating that a selected subset of establishments does not drive our findings. Similarly, studies suggest that identifying businesses with the most similar characteristics is crucial to addressing endogeneity concerns in causal inference studies. To this end, our formal analysis categorizes businesses based on the 10th percentile of pre-treatment visitation counts, creating 10 groups aligned with the distribution of pre-treatment visitation data. By further refining the thresholds, particularly for smaller size bins, and combining lower groups (e.g., merging the 0–50 and 50–100 bins into a single 0–100 group), we repeated the matching process. The results $(\beta = 27.76,\ \text{p-value} < 0.001)$, shown in Table \ref{tab:my_table}, reveal that our findings are consistent across varying categorizations and not driven by specific cutoff choices. The potential influence of estimating the impacts of an EVCS within a comparably narrow radius and considering the possibility of utilizing an EVCS during weekdays may influence our treatment effect estimates. To address both concerns, we repeat the main analysis in NYC while implementing the same matching procedure but with two adjustments: 1) using a wider treatment radius of 750 meters, and 2) considering only weekend visits. When expanding the treatment radius to 750 meters, we continue to observe statistically significant effects $(\beta = 20.35,\ \text{p-value} < 0.01)$, although the effect sizes are smaller, revealing that our findings are not driven by the decision to focus on a narrow radius of treatment. On the other hand, focusing on weekend visits reveals systematically larger effects $(\beta = 32.26,\ \text{p-value} < 0.001)$, suggesting that EVCSs are more beneficial for longer, recreational trips rather than shorter, daily commutes. Taken together, these results suggest that the positive and significant causal effects are consistent against the choice of decision parameters selected in the experiment.

\begin{table}
\centering
\caption{Impact of EV Charger installations on customer count across five boroughs in NYC: comparing baseline model results with robustness checks}
\label{tab:my_table}
\begin{tabular}{ll l l l l}
\hline
  & \textbf{\textit{Baseline}} & \textbf{\textit{Dropping}} \textbf{\textit{the}} \textbf{\textit{closest}} & \textbf{\textit{Different pre-visits}} & \textbf{\textit{Wider treatment radius}} & \textbf{\textit{Only considering }} \\
\multirow{2}{*}{}  &  \textbf{\textit{model}} &  \textbf{\textit{establishment}} & \textbf{\textit{size bin}} & \textbf{\textit{(within 750 meters)}} & \textbf{\textit{weekend visits}} \\
\hline
\multirow{3}{*}{Treatment effect} & $28.52^{\ast\ast\ast}$ & $24.50^{\ast\ast\ast}$ & $27.76^{\ast\ast\ast}$ & $20.35^{\ast\ast}$ & $32.26^{\ast\ast\ast}$\\
\cline{2-6}
  & (4.59) & (4.29) & (5.25) & (7.84) & (5.63) \\
\cline{2-6}
  & $p<0.001$ & $p<0.001$ & $p<0.001$ & $p<0.01$ & $p<0.001$ \\
\hline
Fixed-effects: &   &   &   &   &   \\
% \hline
    Individual POI & Yes & Yes & Yes & Yes & Yes \\
Year & Yes & Yes & Yes & Yes & Yes \\
\hline
Observations & 6932 & 6240 & 7792 & 9712 & 6932 \\
R\textsuperscript{2}  & 0.969 & 0.971 & 0.968 & 0.954 & 0.957 \\
\hline
\end{tabular}
Note. Clustered (at the POI level) standard-errors reported in parentheses, and \textit{p}-values from two-sided \textit{t}-tests are listed under standard errors. The dependent variable is the number of customers. ***p <0.001; **p <0.01; *p <0.05.
\end{table}

The key assumption of the study is that differences in customer foot traffic to businesses are determined by the introduction of an EVCS rather than the impacts of spurious trends. To validate this hypothesis and identify any treatment effects that may arise by chance, we conduct a placebo test for the staggered adoption approach by randomly shuffling the opening dates of EVCSs, as illustrated in Figure \ref{Figure 2}c. Using this random assignment process, ten placebo datasets were generated. After selecting control POIs using the staggered adoption approach, we perform the staggered DID analysis for each placebo dataset to ensure that the estimated treatment effects are not biased by chance. The placebo test, shown by the red line in Figure \ref{Figure 2}d, yields insignificant effects in both NYC $(\beta = 8.44,\ \text{p-value} = 0.112)$ and California Bay Area $(\beta = 1.28,\ \text{p-value} = 0.066)$, indicating that no spurious treatment effects arise when treatment dates are randomly assigned.

\begin{figure}
    \centering
    \includegraphics[width=1\linewidth]{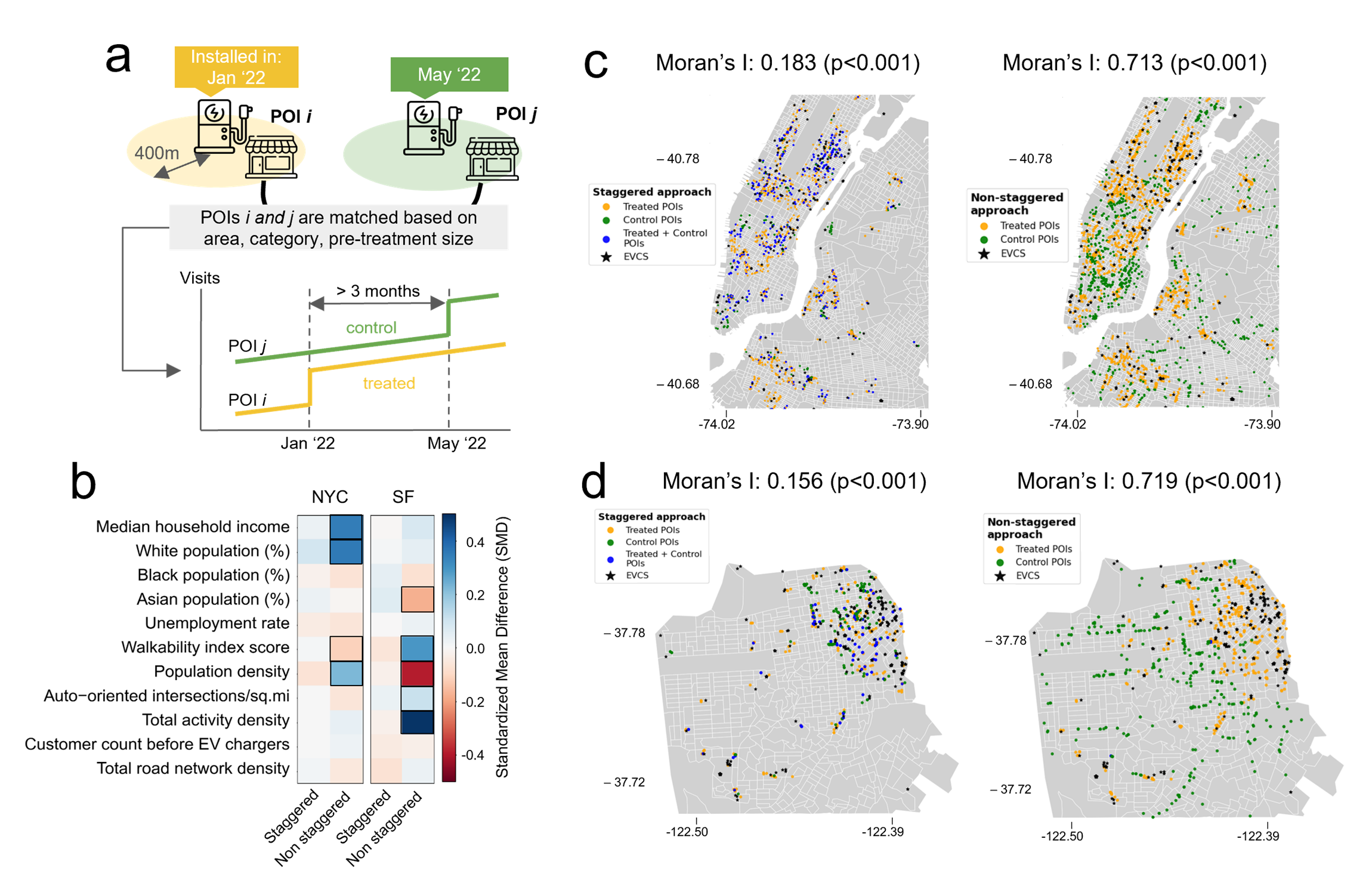}
    \caption{\textbf{Staggered EV charging station deployment enables POI matching for robust causal inference.} a) Selection criteria for control and treatment POIs, where POIs \textit{i} and \textit{j} are situated within the same borough and business category, with similar pre-visitation count. These POIs form a structured comparison set, each impacted by an EVCS at different times. The staggered adoption strategy ensures temporal alignment in treatment, with minimum of three months between successive EVCS installations. The green line denotes POI \textit{j}, acting as a control for treated POI \textit{i}, subsequently impacted by an EVCS in a later date. b) Covariate balance between treatment and control groups, derived using the staggered adoption approach and non-staggered adoption approach in Manhattan New York City (NYC)(left) and San Francisco (right). In the non-staggered approach, covariates are imbalanced when zooming into the downtown areas, with imbalanced covariates highlighted by black borders. This imbalance is not observed in staggered matching pairs, where covariates are balanced across both urban settings. Spatial and structural properties of matched POI pairs, categorized by treatment type and distance proximity, obtained using the staggered adoption and non-staggered matching approaches, are shown for two different urban settings: Manhattan, NYC (c), and San Francisco (d). The yellow points represent treated POIs, the green points indicate control POIs, and the blue points (in the staggered approach) denote POIs initially treated but later used as controls in subsequent time periods. The Moran’s I statistics confirm greater spatial mixing between treated and control POIs under the staggered approach (Manhattan, NYC: 0.183; San Francisco: 0.156) compared to the non-staggered approach (Manhattan, NYC: 0.713; San Francisco: 0.719; all p-values < 0.001). Results suggest that staggered POI matching effectively enhances spatial overlap between treated and control groups, especially in high-density EVCS settings where non-staggered matching often produces spatial segregation. Maps were produced in Python using the TIGER shapefiles from the U.S. Census Bureau\cite{UnitedStatesCensus2022}.}
    \label{Figure 1}  
\end{figure}

\begin{figure}
    \centering
    \includegraphics[width=1\linewidth]{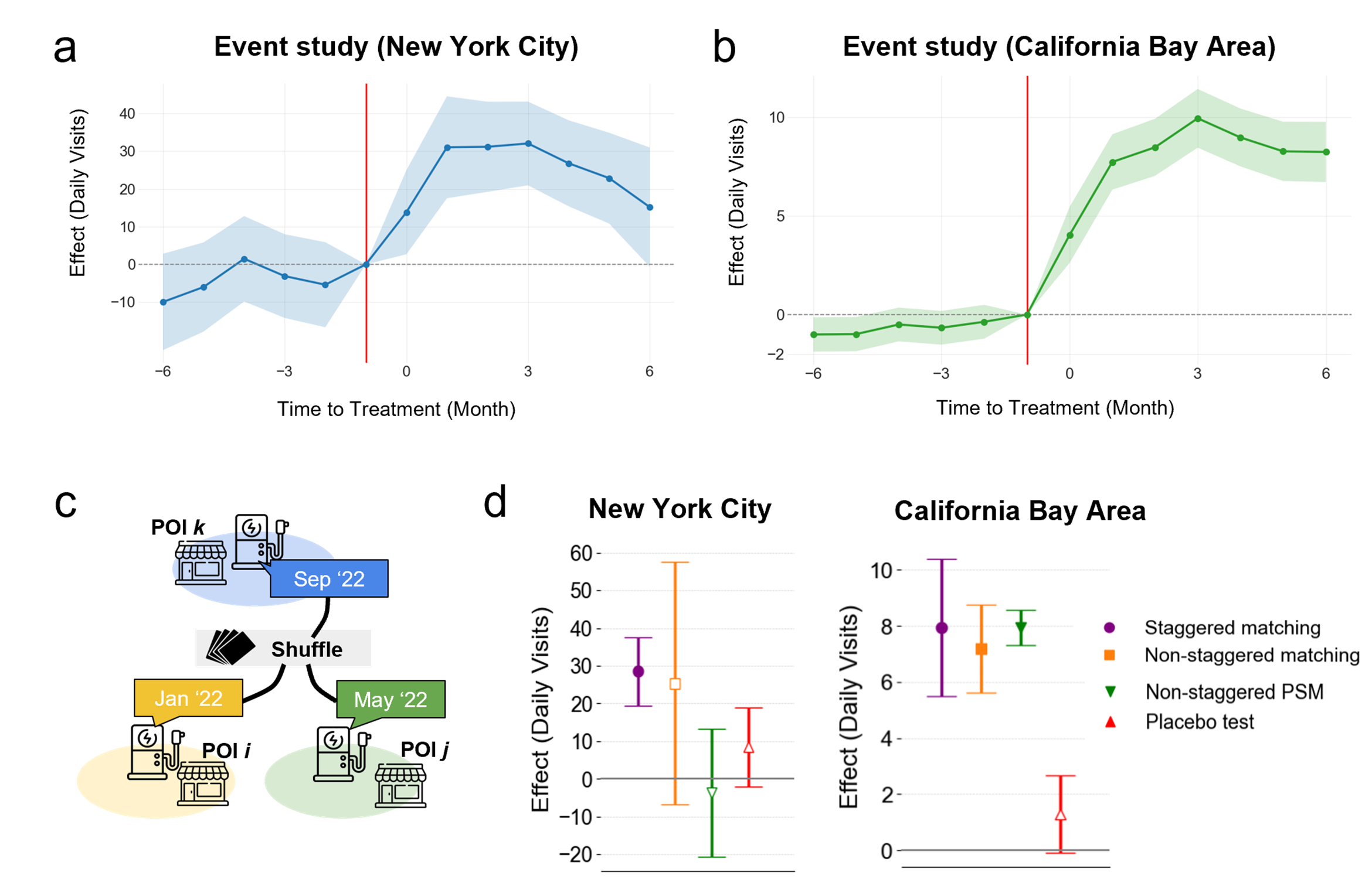}
    \caption{\textbf{Temporal variations in treatment effects and placebo tests for the impact of EV charger installations.} Time variation of the treatment effects on customer counts at surrounding POIs analyzed through event studies in NYC (a) and California Bay Area (b), with the center of shaded error bands representing point estimates and error bands representing the 95\% confidence intervals. The red vertical line represents the baseline period, which is one month before the treatment. Effects were statistically insignificant prior to EVCS installation across both urban settings, supporting the parallel trend assumption of the DID strategy, but turned significantly positive afterward, highlighting EVCS installations increase visits to neighboring businesses. c) The placebo test for the staggered adoption approach is performed by randomly shuffling the opening dates of EVCSs and assigning a hypothetical placebo opening time point for each EVCS.
    % This assumes that by assigning a false EVCS opening date, it is possible to assess whether the estimated treatment effect in the actual treatment group is biased. 
    d) Difference-in-Differences (DID) estimates for the impact of installing a single EVCS on the increase in customer counts at surrounding POIs in NYC (left) and California Bay Area (right). The significant DID estimates from the staggered adoption approach (purple bar line) highlight its robustness in capturing the causal impacts of EVCS installations, particularly in EVCS dense areas of NYC, where both non-staggered matching (orange bar line) and propensity score matching (green bar line) fail to address biases arising from limited spatial overlap. In California Bay Area, an area with moderate urban density and high EV adoption rates, both staggered and non-staggered approaches yield positive results, supporting the robustness of the staggered adoption method across varying urban densities. The placebo test results (red bar line) demonstrate no significant effects, reinforcing the validity of the parallel trends assumption followed in the staggered matching approach. Error bars represent 95\% confidence intervals.}
    \label{Figure 2}
\end{figure}

Figure \ref{Figure 2}d summarizes the estimated impact of a single EVCS installation on nearby customer visits, comparing the staggered adoption design with the corresponding non-staggered estimates for both NYC and California Bay Area. The staggered adoption approach (purple line) shows that the installation of a single EVCS increases customer visits by approximately $\sim 28$ in NYC $(\beta = 28.52, \text{p-value} < 0.001)$ and by $\sim 7$ in California Bay Area $(\beta = 7.93, \text{p-value} < 0.001)$. In contrast, the non-staggered matching (orange line) and non-staggered PSM (green line) approaches perform poorly in NYC, yielding imprecise and statistically insignificant estimates $(\beta = 24.93,\ \text{p-value} > 0.05)$ and $(\beta = -3.63,\ \text{p-value} > 0.05)$, respectively. These results highlight the limitations of non-staggered designs in dense EVCS environments, where residual confounding and spatial heterogeneity hinder credible causal estimation. This limitation aligns with the covariate imbalances visible in Figure \ref{Figure 1}b, where treated and control POIs in Manhattan show limited spatial and contextual overlap. By comparison, both non-staggered approaches perform substantially better in California Bay Area, where EVCS density is lower and spatial overlap is greater. The non-staggered matching and PSM methods yield estimates of $(\beta = 7.17,\ \text{p-value} < 0.001)$ and $(\beta = 7.94,\ \text{p-value} < 0.001)$, respectively, which are closely aligned with the staggered estimate. Taken together, the results illustrate that while both staggered and non-staggered approaches can perform well in less dense environments, only the staggered framework reliably isolates the causal effect of EVCS installations in dense urban settings such as in NYC (see Methods for more details on the staggered DID model). The full regression tables are provided in Supplementary Tables S10 and S11.

\subsection*{Heterogeneous causal effects across POI types}

\begin{figure}
        \centering
        \includegraphics[width=1\linewidth]{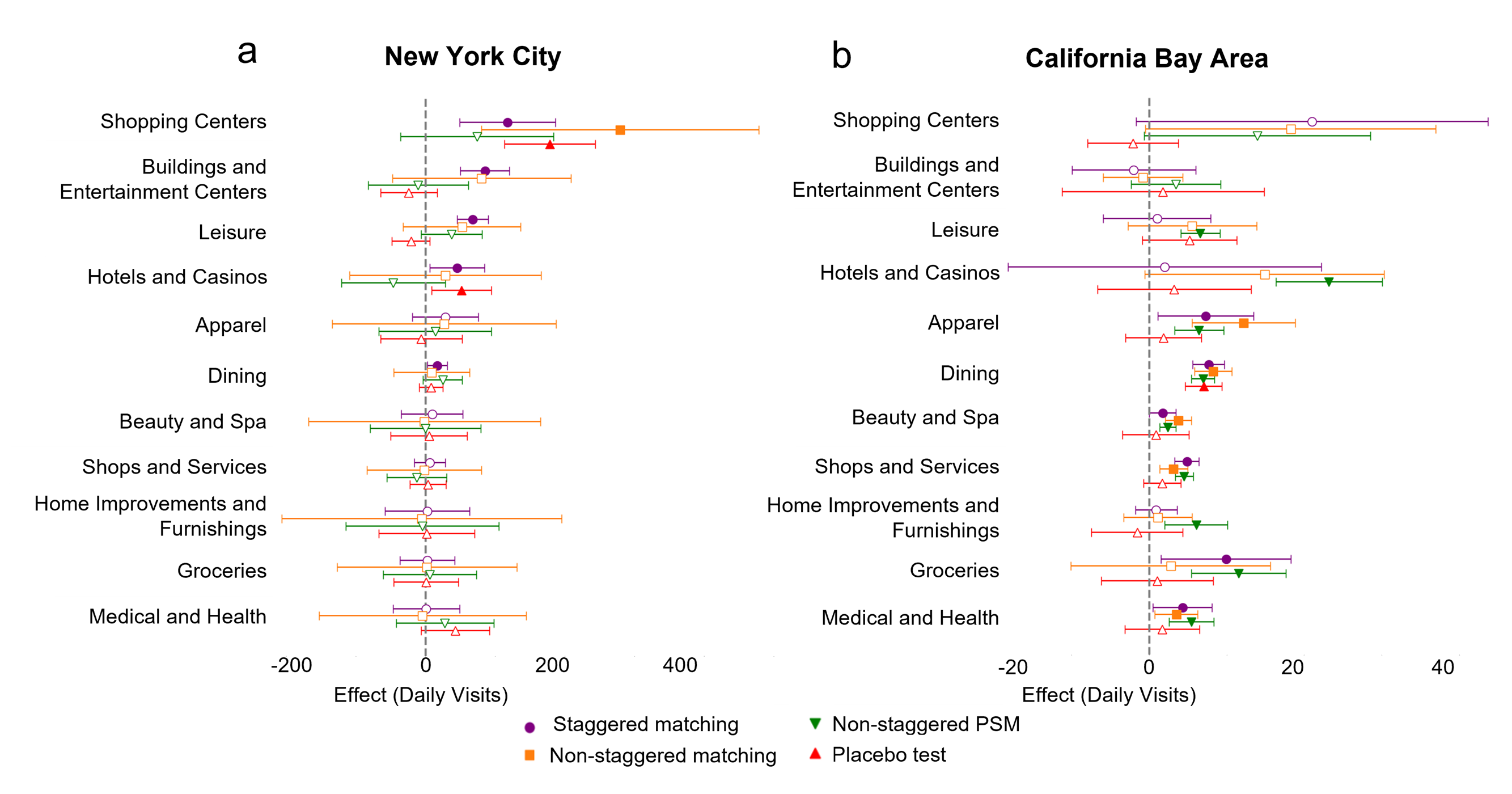}
        \caption{\textbf{Heterogeneous causal effects of EVCS placement across POI types.} These figures illustrate the causal impacts of EVCS installations on customer counts in nearby businesses, stratified by POI category, in NYC (a) and California Bay Area (b). In NYC (a), treatment effects (purple bar lines) are largest for Shopping Centers $(\beta = 117.83,\ \text{p-value} < 0.001)$, Buildings and Entertainment Centers $(\beta = 85.73, \text{p-value} < 0.001)$, Leisure $(\beta = 67.76,\ \text{p-value} < 0.001)$, Hotels and Casinos $(\beta = 45.81,\ \text{p-value} < 0.05)$ and Dining $(\beta = 14.94,\ \text{p-value} < 0.05)$, suggesting that newly attracted customers are primarily recreational or exploratory visitors rather than regular patrons. Placebo tests (red bar lines) show significant effects for Shopping Centers $(\beta = 179.11, p < 0.05)$ and Hotels and Casinos $(\beta = 51.79,\ \text{p-value} < 0.05)$, indicating potential data imbalances. b) In contrast, California Bay Area (b) exhibits distinct behavioral preferences among EV drivers, with stronger effects for Groceries $(\beta = 9.94,\ \text{p-value} < 0.05)$, Dining $(\beta = 7.70,\ \text{p-value} < 0.001)$, Apparel $(\beta = 7.32,\ \text{p-value} < 0.05)$, Shops and Services $(\beta = 4.90,\ \text{p-value} < 0.001)$, Medical and Health $(\beta = 4.33,\ \text{p-value} < 0.05)$, and Beauty and Spa $(\beta = 1.77,\ \text{p-value} < 0.05)$, reflecting the city’s moderate urban density and different charging use contexts compared to NYC’s dominance of recreational POIs. In both plots, error bars represent 95\% confidence intervals.}
    \label{Figure 3}
\end{figure}

The staggered matching approach consistently shows significantly positive effects of newly installed EVCS on customer counts at surrounding POIs across both NYC and California Bay Area (Figure \ref{Figure 2}d). Next, we determine whether these effects are homogeneous across different POI categories. To measure this heterogeneity, we employ the following model specification to estimate treatment effects stratified by POI types: 
$Y_{it} = D_i \times T_t \times \sum_{c=1}^{11} \beta_c \text{Cat}_{ict} + u_i + \omega_{\text{year}} + \epsilon_{it}$. Here, $\text{Cat}_{ict}$ represents customer visits for POI Category \textit{c} at POI \textit{i} during time period \textit{t}, where \textit{c} can take on one of eleven possibilities: Apparel, Beauty and Spa, Buildings and Entertainment Centers, Dining, Groceries, Home Improvements and Furnishings, Hotels and Casinos, Leisure, Medical and Health, Shopping Centers, and Shops and Services. As shown in Figure \ref{Figure 3}a, the purple bar lines show the treatment effects using staggered adoption approach, where Shopping Centers $(\beta = 117.83,\ \text{p-value} < 0.001)$, Buildings and Entertainment Centers $(\beta = 85.73,\ \text{p-value} < 0.001)$, Leisure $(\beta = 67.76, \text{p-value} < 0.001)$, Hotels and Casinos $(\beta = 45.81,\ \text{p-value} < 0.05)$ and Dining $(\beta = 14.94,\ \text{p-value} < 0.05)$ are positive and significant, indicating that the presence of EV chargers may influence consumer behavior, encouraging visits to convenience type stores, restaurants, and recreational/rest areas during charging times. This suggests that newly attracted customers are more likely to be recreational or exploratory visitors than regular and loyal customers in NYC. In contrast, in California Bay Area (Figure \ref{Figure 3}b), the staggered approach reveals distinct behavioral preferences among EV drivers, with stronger effects observed for Groceries $(\beta = 9.94,\ \text{p-value} < 0.05)$, Dining $(\beta = 7.70,\ \text{p-value} < 0.001)$, Apparel $(\beta = 7.32,\ \text{p-value} < 0.05)$, Shops and Services $(\beta = 4.90,\ \text{p-value} < 0.001)$, Medical and Health $(\beta = 4.33,\ \text{p-value} < 0.05)$, and Beauty and Spa $(\beta = 1.77,\ \text{p-value} < 0.05)$. These patterns indicate that EV drivers disproportionately frequent routine grocery destinations, pharmacies, and cafés, reflecting San Francisco’s moderate urban density and distinct charging-related activity patterns, which contrast with the recreational POI driven effects observed in NYC.

The non-staggered matching approach in NYC, shown in orange bar lines, reveals significant effects only for Shopping Centers $(\beta = 279.83,\ \text{p-value} < 0.05)$, while the non-staggered PSM approach (green bar lines) fails to capture robust treatment effects for any POI category. This lack of significance highlights the reduced sensitivity and potential covariate imbalance of control groups when temporal information is ignored, particularly in high density EVCS environments. However, in California Bay Area (Figure \ref{Figure 3}b), both non-staggered approaches yield significant effects for categories such as Dining, Apparel, Medical and Health, and Beauty and Spa, and these estimates closely align with those obtained from the staggered approach. This suggests that the staggered POI matching approach achieves greater spatial overlap specially in dense urban areas, reducing bias and providing more consistent results across cities. The placebo test results for treatment effects across various POI categories, shown in red bar lines, indicate significant results in New York City for Shopping Centers $(\beta = 179.11,\ \text{p-value} < 0.05)$ and Hotels and Casinos $(\beta = 51.79,\ \text{p-value} < 0.05)$, however, could be due to the bias in the dataset (i.e., there are fewer matched pairs in these two categories, which may reflect some underlying imbalance in the data). The full DID estimations stratify by POI category across both cities are shown in Supplementary Tables S12 and S13. 

\subsection*{Spatial variations in treatment effects and heterogeneity by across clientele}
\begin{figure}
    \centering
    \includegraphics[width=1\linewidth]{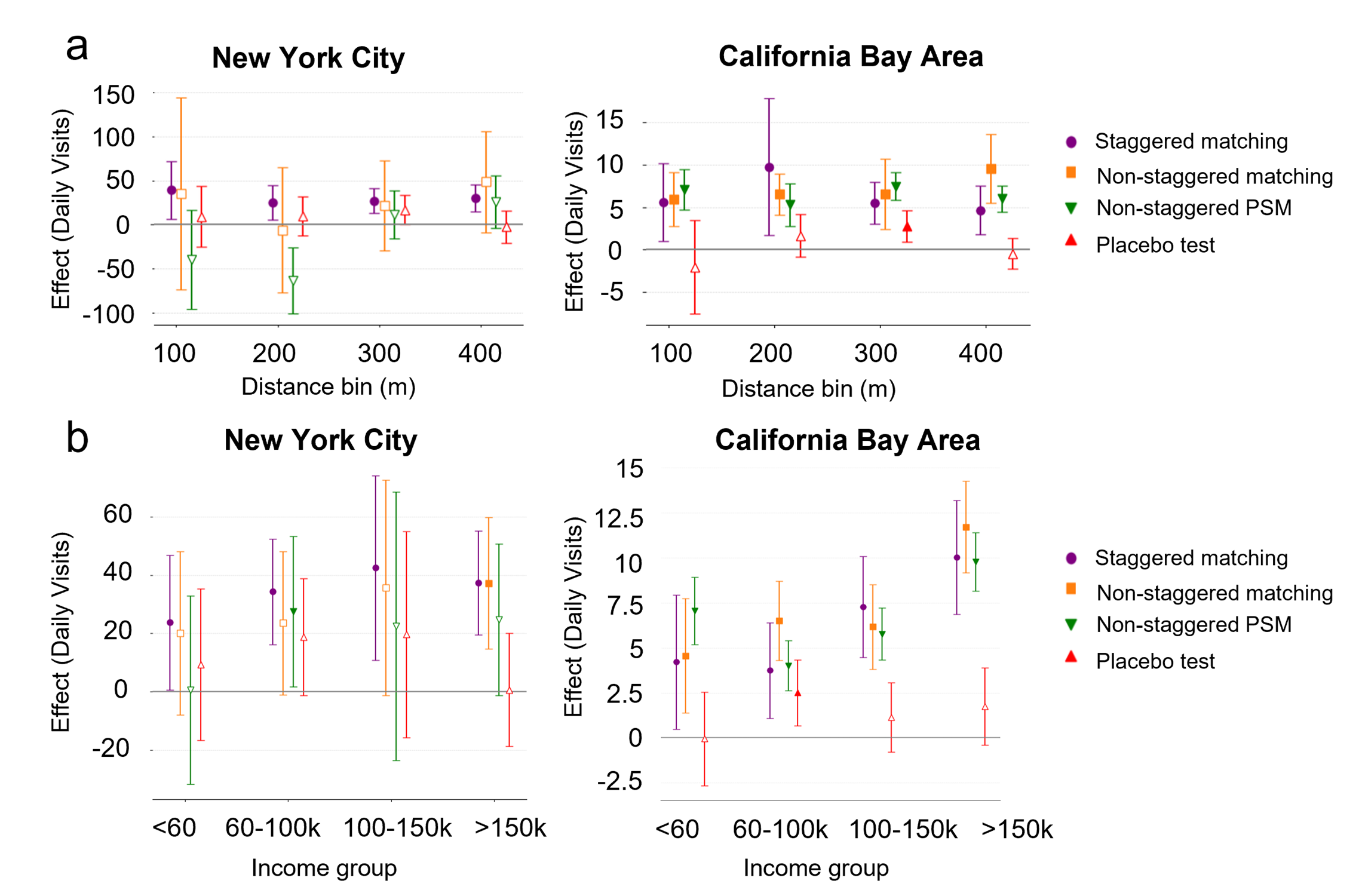}
    \caption{\textbf{Spatial variations in treatment effects and heterogeneous causal effects across clientele.}
     a) Variations in the treatment effects of EVCS installations on customer counts based on distance between the POIs and nearby EVCSs, in NYC (left) and California Bay Area (right). The effects (purple bar lines) estimated using the staggered adoption approach are more pronounced for POIs within 100 meters of an EVCS in NYC, where customer counts increase by $\sim 39$, whereas in California Bay Area, POIs located 100--200 meters from an EVCS attract more customers, with counts increasing by $\sim 10$. Compared to non-staggered matching approaches and the placebo test, distance-varying treatment effects reveal bias in high density EVCS settings in NYC, whereas in California Bay Area, with a moderate urban density, the results are consistent across approaches. Thus, these results provide nuanced insights into how we can generate more reliable treated and control POI pairs using the staggered adoption approach, further emphasizing its robustness across spatial categories. b) The heterogeneous effect of an additional EVCS on customer counts among different income groups at surrounding businesses, in NYC (left) and California Bay Area (right). We find that EVCS installations attract customers from mid high-income households $(\$100-150k,  \beta = 42.43, \text{p-value} < 0.001)$ most strongly in NYC, whereas in California Bay Area, businesses serving high income households $(>\$150k, \beta = 10.02, \text{p-value} < 0.001)$ experience the largest positive effects. Results suggest that businesses serving higher-income customers experience greater impacts, aligning with the assumption that EV drivers are more likely to belong to higher-income groups. Insignificant results from the non-staggered matching approaches and placebo test across income segments in NYC further confirm the integrity of the staggered adoption approach, highlighting the risk of bias due to limited spatial overlap in high-density EVCS settings. In contrast, California Bay Area, where urban density is moderate, non-staggered approaches yield consistent and significant effects for higher income segments, illustrating that the reliability of non-staggered methods depends on spatial context. In both plots, error bars represent 95\% confidence intervals.}  
    \label{Figure 4}
\end{figure}

For effective planning and enhancing the resilience of urban development, it is important to determine the extent convenience of access significantly influences consumer choices. We first categorize the distances between treatment POIs and their nearby EVCS into four distance bins, ranging from 0 to 400m with 100m increments. Figure \ref{Figure 4}a illustrates the variations in treatment effects of EVCS installations on customer counts across these distance bins in NYC and California Bay Area, respectively (see Methods for more details on the distance-varying treatment effect). The purple bar lines represent the cumulative effects of adding one EVCS on customer counts using the staggered adoption approach, with the closest distance bin (0–100m) showing the most substantial impact. Specifically, within this proximity, the treatment effect leads to a significant increase of $\sim 39$ $(\text{p-value} < 0.05)$ in customer counts in NYC, whereas in California Bay Area, POIs located 100--200 meters from an EVCS experience an increase of $\sim 10$ customers. This pattern reflects the contrasting spatial structures of the two cities: the high density and mixed-use clustering in NYC facilitate strong, localized spillover effects within short walking distances, consistent with urban planning theories on agglomeration and pedestrian accessibility. In contrast, SCalifornia Bay Area’s lower density, more spatially dispersed layout distributes customer flows over broader areas, resulting in smaller but still positive effects at slightly larger distances from EVCS locations. However, as the distance from the EVCS increases to the range of 300--400m, the magnitude of the treatment effect diminishes, resulting in a smaller yet still significant increase of $\sim 30$ $(\text{p-value} < 0.001)$ customer visits in NYC and $\sim 4$ $(\text{p-value} < 0.001)$ customer visits in California Bay Area. These results advocate for the incorporation of consumer preferences in accessibility, improving the ability to capture observed movement patterns and highlighting the possibility that EVCSs are systematically placed in locations where they are more likely to benefit co-located businesses, providing a more robust framework for strategic EVCS placements\cite{Tsou2013,Yang2021}. 

The observed variations in treatment effects using the two non-staggered matching approaches, shown as orange lines (non-staggered matching) and green lines (non-staggered PSM) in Figure \ref{Figure 4}a, highlight concerns about the reliability of non-staggered designs in high density EVCS settings. In California Bay Area, the results are broadly consistent with the patterns identified using the staggered adoption approach, whereas in NYC, substantial inconsistencies remain evident. We also find that the non-staggered matching results are counterintuitive from an urban planning perspective, as the farthest distance bin (300–400 meters) exhibits the largest estimated effect in both urban settings: $(\beta = 48.52,\ \text{p-value} > 0.05)$ in NYC and $(\beta = 9.56,\ \text{p-value} > 0.05)$ in California Bay Area. The unexpected magnitude and direction of effects in more distant locations emphasize the limitations of non-staggered matching strategy in accurately capturing the treatment effects. The red bar lines in Figure \ref{Figure 4}a represent the placebo test results for treatment effects across the spatial dimension. The insignificant results for each distinct distance bin indicate that the dataset generated using the staggered matching approach is not biased for distance categories. These mixed results of distance-varying treatment effects provide nuanced insights into how we can generate more reliable treated and control POI pairs using the staggered adoption approach, highlighting its robustness across spatial categories. Supplementary Tables S14 and S15 show the full regression results. 

We further investigate the impact of EVCS installations on customer counts across various income groups. The average income of the visitors during the 3-month period before the placement of the EVCS was estimated for each POI, by analyzing which CBGs the visitors came from. To explore potential heterogeneous treatment effects based on visitor income, we categorize visitors into four income groups based on their annual household income: $<\$60k$, $\$60$–$100k$, $\$100$–$150k$, and $>\$150k$. For each POI, we identify the income group that constitutes the largest proportion of visitors, thereby capturing the primary demographic profile of its customer base (see the Methods section for details). The results of the staggered adoption approach, shown by the purple bar lines in Figure \ref{Figure 4}b, reveal significant and positive effects of EVCS installations on customer counts across all income groups in both urban settings. This suggests that government efforts to promote equity in both EV adoption and EVCS accessibility may have played a role in enabling low-income individuals to acquire and charge EVs\cite{Jenn2020}. However, we find that EVCS installations are most effective in attracting customers from mid high-income households $(\$100-150k,\ \beta = 42.43,\ \text{p-value} < 0.001)$ in NYC, whereas in California Bay Area, businesses serving high-income households $(>\$150k,\ \beta = 10.02,\ \text{p-value} < 0.001)$ experience the largest positive effects. These patterns reflect the broader trend that EV drivers predominantly belong to higher income demographics. In NYC, EV adoption remains relatively low, at just around 1\% of light duty vehicles according to recent estimates, despite a generally affluent population, suggesting that mid high income households represent the threshold segment for EV uptake\cite{HRAAdvisors2023}. Meanwhile, in the California Bay Area, EV penetration is substantially higher, around $\sim 9\%$ in San Francisco, with a median household income of approximately \$122,000 \cite{MTC2025}. EV ownership is concentrated in the highest-income ZIP codes, where residents are more likely to use nearby charging infrastructure. Thus, while mid high income households in NYC represent the threshold segment for EV uptake, in California Bay Area, the largest effects are driven by the very highest income households, where EV adoption is already relatively mature.

As shown by the orange bar lines (non-staggered matching) and green bar lines (non-staggered PSM) in Figure \ref{Figure 4}b, the observed variations in treatment effects across income groups reveal bias due to limited spatial overlap in high EVCS density areas. In NYC, only the highest income category $(\beta = 37.20,\ \text{p-value} < 0.001)$ shows a significant effect, whereas in California Bay Area, the results are broadly consistent with the patterns identified using the staggered adoption approach. The placebo test results for treatment effects, shown in red bar lines, capture unbiased sample characteristics. These results provide confidence in the integrity of the staggered adoption approach, lending support to the original hypothesis. Supplementary Tables S16 and S17 show the full regression comparisons.

\section*{Discussion}

Estimating the causal economic impact of EVCSs on nearby businesses presents a complex challenge, particularly in dense urban environments where placement decisions are rarely random. EVCSs are often deployed in areas already characterized by high foot traffic and commercial activity, and this strategic deployment introduces endogeneity concerns that complicate causal inference. Determining whether EVCS installations actually lead to increased customer visitation and spending at proximate POIs is essential, not only to justify public and private investments but also to inform equitable and data-driven infrastructure planning. This challenge is especially acute in urban areas, where EVCS density is highest and future deployments are expected to intensify. As EVCS placement decisions increasingly shape the built environment, there is a growing need for methods that can produce credible causal estimates in complex urban contexts \cite{Zheng2024,Esmaili2024,Liang2023}. Moreover, there remains a critical gap in developing a transferable causal inference framework that can be applied across national settings, enabling consistent evaluation of EV infrastructure rollouts in diverse urban environments.

In this context, this study makes four key contributions to supporting EV network operators’ claims that strategically placing an EVCS can provide local businesses with access to a new, desirable consumer audience \cite{Roy2022,Ito2024,Esmaili2024}. First, we introduce a staggered adoption strategy to estimate the impact of EVCS installations on nearby businesses in two distinct urban settings, by better accounting for the timing and location of installations. Compared to prior methods \cite{Zheng2024,Babar2024}, the staggered approach generates more geographically and temporally comparable matched groups and reduces potential biases from unobserved confounders. This is particularly important in urban contexts, where EVCS deployments are often concentrated in commercially vibrant areas, making it difficult to disentangle true treatment effects from underlying location characteristics. By leveraging variation in EVCS rollout timing, our methodology generates more comparable treatment and control groups, increasing spatial overlap in dense areas and thereby enhancing the validity of causal estimates. Empirically, we find that this approach produces significantly stronger effects than non-staggered adoption methods in NYC, which exhibits high EVCS density, and in California Bay Area, characterized by moderate urban density and high EV adoption rates, highlighting a new methodological advancement for urban science research. From a practical perspective, the results support the view that well-planned EVCSs can generate spillover economic benefits for co-located retail businesses by attracting a growing segment of EV-driving consumers. This has important implications as EV adoption accelerates: while there are currently about 20 EVs per public charging outlet nationwide (based on 2024 state-level data), the ratio of gas-powered vehicles to gas pumps is approximately 1600:1, highlighting the untapped potential of EVCS as a consumer attraction tool \cite{Babar2024}. As the EV market expands, these spillover effects are likely to become even more pronounced, positioning EVCS deployment as both a sustainability measure and a local economic development strategy. 

Second, we show that the business impact of EVCS installations varies by establishment type and urban context. In NYC, the largest effects are observed for convenience stores, restaurants, and recreational venues, suggesting that EV charging times create windows of consumer downtime that are more easily converted into visits to businesses offering quick or leisure focused experiences. These visits are largely recreational or exploratory rather than habitual, indicating that new customer acquisition is a key mechanism behind the observed gains. In contrast, in California Bay Area, EV drivers disproportionately visit routine destinations such as grocery stores, pharmacies, and cafés, reflecting the city’s moderate urban density and the more practical, everyday nature of charging behavior. Together, these findings highlight how the local urban and behavioral context shapes the spillover effects of EVCSs. Notably, we find that the cumulative effect of EVCS installations on all nearby businesses is substantially larger than the effect on any individual POI, highlighting the benefits of multi-host EVCS deployment. This presents a compelling case for coordinated strategies where multiple businesses collaborate to co-locate charging infrastructure and jointly benefit from increased consumer traffic.

Third, we demonstrate that a POI with a newly opened EVCS within 400m experiences a significant increase in customer visits. In NYC, the dense, mixed-use urban environment concentrates customer flows within short walking distances, producing strong localized effects for businesses closest (0–100m) to the charging sites with an increase of 39 visits $({p-value} < 0.001)$. In contrast, in California Bay Area, the moderately dense and more spatially dispersed urban environment distributes customer activity over a wider area, resulting in smaller but still positive spillovers at intermediate distances, with POIs located 100--200 meters from an EVCS experiencing an increase of 10 customers $({p-value} < 0.001)$. Across both cities, the estimated benefit decays steadily with increasing distance from the charging site, indicating that proximity plays a crucial role in realizing economic spillovers from EVCS. Given that EV network operators typically avoid placing redundant chargers in close geographic proximity, this spatial decay suggests a first-mover advantage for businesses located near newly installed chargers. Businesses seeking to benefit from increased foot traffic should act early, as their location relative to the EVCS is likely to determine the magnitude of the return. 

Finally, our results suggest that EVCS installations currently attract more visits from higher-income brackets, reflecting the demographic profile of early EV adopters\cite{Muehlegger2018,Farkas2018}. This disparity underscores the importance of considering demand side factors, particularly consumer demographics, which play a critical role in shaping the economic returns of EVCS investments. While these patterns highlight where current returns may be concentrated, they also point to an emerging challenge: ensuring that the growth of EV infrastructure does not inadvertently reinforce existing inequalities in access and benefit. As EV adoption continues to expand beyond early high-income adopters, it will be essential to monitor how the spatial distribution of EVCS aligns with income patterns and neighborhood characteristics. Future research should explore how EVCS deployment can support equitable outcomes, particularly by encouraging visits to underserved commercial areas and examine whether charging infrastructure (charger types) can play a role in reducing spatial segregation. From a practical perspective, our research highlights the potential for policymakers and businesses to leverage EVCS as a catalyst for local economic growth, particularly in diverse commercial areas. Our findings suggest that EVCS providers can adopt a business model similar to the traditional paradigm of \textquotedblleft gas station--convenience store\textquotedblright, where fuel stations are strategically paired with retail establishments to maximize consumer spending\cite{NACS2019}. Just as gas station owners capitalize on ancillary sales, EVCS operators and local businesses could collaborate to internalize the economic benefits of EVCS installations, transforming charging locations into hubs of commercial activity that attract and retain customers\cite{Arlt2023}.

While our staggered adoption approach offers a novel perspective on mitigating endogeneity concerns by leveraging variations in the timing of EVCS installations, it also lays the foundation for future research endeavors. First, while our analysis leverages robustness checks and placebo tests to support the validity of the parallel trends assumption, future research could build on these findings by incorporating direct economic outcome measures. Second, our current approach uses human mobility as a proxy for business activity, but future studies may enhance precision by integrating data on spending (\$ estimates) in nearby businesses to more directly quantify the economic effects. Additionally, alternative empirical strategies like synthetic control methods or instrumental variable approaches could be explored to further validate the results. Lastly, our study focuses on the immediate impact of EVCS installations on customer foot traffic and business activity; however, businesses may adapt their strategies over time by adjusting their service offerings, marketing approaches, or even relocating in response to changes in consumer behavior. Understanding how businesses evolve in response to the presence of EVCS would require long-term observational data and dynamic treatment effect models. 

\section*{Methods}

\subsection*{Study context and data}
Our information on public electric vehicle charging stations (EVCSs) was obtained from the Department of Energy’s Alternative Fuels Data Center\cite{USDOE2023}. This dataset provides comprehensive details including opening dates, geographical coordinates (longitude and latitude), the number of ports for each type of EV charger (Level 1, Level 2, or DC fast chargers), and accessibility status (public or private) and is currently the most complete publicly available EVCS database in the United States. However, some EVCSs may not be updated in real time within the dataset. For the scope of our analysis, we focus exclusively on public EVCSs opened between January 2019 and January 2024 across the five boroughs of New York City. To assess how our proposed staggered adoption approach operates in urban areas with moderate charging station density and high EV adoption rates, we extend the analysis to California Bay Area with the highest EVCS density measured in stations per square kilometer: Santa Clara (5.987), San Francisco (4.048), San Mateo (2.883), Alameda (2.771), Orange (2.758), and Los Angeles (2.566). These counties represent some of the most EV intensive urban environments in the United States, providing a strong basis for evaluating the generalizability of our method across different but comparable metropolitan contexts. Furthermore, this time frame is crucial because data on customer visits at NYC POIs is only available from January 2019 onward, making any analysis prior to 2019 infeasible. Figure S1 in the Supplementary Information illustrates the selected areas in New York City by showing the number of EVCSs in each borough during the study period, whereas Figure S2 presents the San Francisco county and the number of EVCSs over the same period.

Our data on community sociodemographic data was collected from the American Community Survey 2022 5-year estimates\cite{USCensus2022}, which provide detailed information on over 329 million Americans, including population statistics, median household income, employed population, gender distribution, and ethnic composition at the Census Block Group (CBG) level (refer to Supplementary Table S2 for description of the tables). Additionally, data on auto-oriented intersections per square mile, total activity density, and total road network density at the block group level were sourced from the Environmental Protection Agency’s (EPA) Smart Location Database\cite{USEPA2022}. National Walkability Index score, which measures walkability at the CBG level, was also obtained from the same database\cite{USEPA2022}. 

Data on NYC and California Bay Area wide activity locations were sourced from Placer.ai’s Global POIs Metadata, which includes category and geographical location information for over 10 million unique places across the U.S. as of January 2024. The dataset covers POIs across more than 600 North American Industry Classification System (NAICS) codes and 250 major categories, including ‘Dining’ and ‘Shops and Services’\cite{PlacerAI2024}. For our analysis, we exclusively considered POIs falling within the following eleven categories: (1) apparel; (2) beauty and spa; (3) buildings and entertainment centers; (4) dining; (5) groceries; (6) home improvements and furnishings; (7) hotels and casinos; (8) leisure; (9) medical and health; (10) shopping centers; and (11) shops and services. Mobility pattern data for NYC and California Bay Area was obtained from the Placer API, part of the Placer.ai Data platform\cite{PlacerAPI2024b}. This dataset tracks anonymous mobile device GPS movements and, for each POI, provides insights such as raw visit counts, median dwell times, and visitations per each CBG level. It’s important to note that cash transactions are not captured by the dataset, which presents a limitation of the data. Additional steps were performed to process POIs (refer to Supplementary note 1.1) and analyze the heterogeneity of visitor characteristics (see Supplementary note 1.3).

\subsection*{POIs matching strategies}
EVCS providers strategically deploy charging stations in locations where their impact is expected to be maximized, often selecting areas with high consumer traffic, economic activity, or accessibility advantages\cite{He2016,Ren2019,Liang2023}. As a result, POIs near EVCS installations may inherently differ from those without nearby chargers, introducing potential selection bias in estimating the true effects of EVCS on local businesses. As the deployment of EVCS continues to expand, an increasing number of locations will fall under this EVCS-dense category, making it even more critical to distinguish the causal effects from underlying site characteristics\cite{Liang2023}. To address this, we compared the staggered adoption approach with a non-staggered matching approach and a non-staggered PSM approach, to evaluate whether ignoring the intrinsic characteristics of EVCS site selection introduces bias into the causal inference. These distinct methods create balanced comparison groups, allowing us to evaluate their effectiveness in accurately estimating the impact of EVCS installations. In addition, to examine how our proposed staggered adoption approach functions in urban areas with moderate charging station density and high EV adoption rates, we apply all three matching approaches to California Bay Area with the highest EVCS density. This broader application enables us to evaluate the performance of each method across different urban contexts.

Our staggered adoption approach leverages the staggered rollout of EVCS installations across different retail locations over time. POIs within a 400 m radius of newly installed EVCS are designated as treated and control groups, with a minimum of three months between successive EVCS installations to ensure proper temporal alignment. In the matching process, we select POIs that are located in the same borough or county and belong to the same business category. To ensure comparability, the control POI must be associated with an EVCS different from that linked to the treated POI, and the control station must have been installed at least three months after the treated station's installation date. Among the set of candidate control POIs, we identify the most similar neighbor by comparing their visit counts observed over the 3-6 months leading up to the arrival of the EVCS. Given the computational burden involved in this process (due to the large volume of candidate retailer pairs), we use a greedy nearest neighbor algorithm \cite{ho2007matching}. This matching criteria involves a minimization approach, where the most similar neighbor was selected by minimizing the absolute difference in DID estimates following the parallel trend assumption. More specifically, we define the 'pre-intervention difference' as the difference in average visits between 3-6 months before the intervention and the average visits between 1-3 months before the intervention. For NYC, this approach produces a panel of 2,645 treated establishments, consisting of 1,733 treated POIs and 912 unique POIs that serve as controls until they are later affected by an EVCS installation. For California Bay Area, the resulting panel includes 16,151 treated establishments, made up of 9,690 treated POIs and 6,461 unique POIs that function as controls prior to receiving an EVCS at a later point in time.

In both non-staggered matching approaches, untreated POIs were designated as those located outside the 400m radius of any EVCS opened during the same period. The non-staggered matching approach follows a two-step method inspired by the staggered adoption approach, to match treated POIs with control POIs that share similar characteristics. First, we performed an exact match based on the same borough \textit{b} and POI category \textit{c}, ensuring that each treated POI is paired with a control POI, such that $i, j \in \{ \text{POIs}(b, c) \}$, followed by the nearest-neighbor algorithm to minimize the absolute difference in DID estimates. For NYC, this procedure produces a matched panel of 3,553 unique establishments, of which 2,066 ultimately receive treatment. For California Bay Area, the procedure yields a matched panel of 18,870 unique establishments, including 10,601 that are eventually treated. In contrast, the non-staggered PSM approach pairs treatment POIs with control POIs of the same category and similar characteristics. These characteristics include: (1) built environment factors (such as population density, auto-oriented road/total road network density, and walkability), which reflect accessibility and urban activity, as EVCS are often installed in densely populated, vehicle-accessible areas; (2) sociodemographic variables (such as income, unemployment, and racial composition), which account for economic conditions and EV adoption potential; and (3) POI-level variables (such as pre-installation customer counts), which indicate areas with high consumer activity and are attractive for EVCS installation. First, we employed logistic regression to regress the treatment variable on these covariates using the \textquotedblleft statsmodels.formula.api.logit\textquotedblright  function in Python, which allowed us to estimate the propensity scores for the covariates. During the matching process, an exact match based on POI categories was first performed to ensure that each treated POI was paired with a control POI from the same category. Then, nearest-neighbor PSM was applied using the \textquotedblleft sklearn.neighbors.KNeighborsRegressor\textquotedblright function in Scikit-learn to match treated POIs with control POIs that shared similar characteristics. For NYC, this procedure resulted in a matched panel of 2,954 unique establishments, including 1,872 that eventually receive treatment. For California Bay Area, the procedure produced a matched panel of 10,812 unique establishments, of which 8,640 eventually become treated.

\subsection*{Staggered difference-in-differences (DID)}
Following the selection of control POIs using the three approaches, we incorporate the matched control POIs alongside the treated POIs and proceed with the staggered DID analysis. The model specification is defined as follows: \begin{equation}
Y_{it} = \beta D_i \times T_t + u_i + \omega_{\text{year}} + \epsilon_{it}\end{equation} In this equation, $Y_{it}$ represents the outcome variable, which is the number of customers at POI \textit{i} during time period \textit{t}. $\beta$ represents the treatment effect, quantifying the impact of adding a new EVCS on the number of customers at the respective POI. The term $D_{i}$ is a binary treatment indicator that takes on the value of 1 for treated establishments when an EVCS opens for service nearby. Similarly, $T_{t}$ is a binary time indicator that shifts from zero to one in the post-treatment period. $u_i$ denotes the POI fixed effects. $\omega_{\text{year}}$ represents the year fixed effects. $\epsilon_{it}$ represents the error term. 

\subsection*{Event study analysis}
To assess the validity of the parallel trend assumption between POIs with nearby EVCSs and those without, and to examine the temporal evolution of the treatment effect, we conduct an event study analysis. The presence of a parallel trend implies that, in the absence of treatment, the difference between the treatment and control groups remains constant over time, strengthening the credibility of our staggered DID approach. The specification of the event study model is given as follows: \begin{equation}
Y_{it} = \sum_{k \neq -1} \beta_k \left( D_i \times 1\{event\_time = k\} \right) + u_i + \omega_{\text{year}} + \varepsilon_{it}
\end{equation} In this specification, $\beta_{k}$ captures how the treatment effect varies over time relative to the EVCS installation, allowing us to quantify the impact of a new EVCS on customer visits at each POI in month ${k}$ relative to the event. The term $D_{i}$ is a binary indicator equal to 1 if POI ${i}$ is treated, meaning that an EVCS has opened nearby, and 0 otherwise. The indicator $\mathbf{1}\{ \text{event\_time} = k \}$ identifies the specific month ${k}$ relative to treatment, so the interaction $\left( D_i \times 1\{event\_time = k\} \right)$ isolates the effect of treatment at each relative time period. 

\subsection*{Distance-varying treatment effect}
In addition to estimating the average treatment effect, we examine how the treatment effect varies based on the distance between POIs and their adjacent EVCSs. This analysis allows us to capture potential spatial heterogeneity in the impact of EVCS installations. To numerically characterize this dynamic effect influenced by distance, we present the model formulation as follows:\begin{equation}Y_{it} = D_i \times  T_t \times \sum_{d=1}^{4} \beta_d \text{Dis}_{idt} + u_i + \omega_{\text{year}} + \epsilon_{it}\end{equation} where, $\text{Dis}_{idt}$ is a binary dummy variable that takes the value of 1 if an EVCS was opened during the study period time \textit{t} and that fall within distance bin d of POI \textit{i}. We utilized four distance bins spanning from 0 to 400m, with 100m increments. Therefore, $\beta_{d}$ represents the treatment effect specific to that particular distance bin d.

\subsection*{Treatment effects by POI types}
Next, we determine whether these effects are homogeneous across different POI categories. To measure this heterogeneity, we employ the following model specification to estimate treatment effects stratified by POI types: 
\begin{equation}Y_{it} = D_i \times T_t \times \sum_{c=1}^{11} \beta_c \text{Cat}_{ict} + u_i + \omega_{\text{year}} + \epsilon_{it}\end{equation} here, $\text{Cat}_{ict}$ represents customer visits for POI Category \textit{c} at POI \textit{i} during time period \textit{t}, where \textit{\textbf{c}} can take on one of eleven possibilities: Apparel, Beauty and Spa, Buildings and Entertainment Centers, Dining, Groceries, Home Improvements and Furnishings, Hotels and Casinos, Leisure, Medical and Health, Shopping Centers, and Shops and Services. $\beta_{c}$ thus capture the treatment effects associated with each POI type. 

\subsection*{Treatment effects on populations from different income groups}
We also explore the impact of EVCS installations on customer counts across various income groups. Using the "visits by CBG reports" provided by the Placer API, we first obtain the 3-month average visits during the pre-treatment period for each POI, segmented by each CBG. To explore potential heterogeneous treatment effects based on visitor income, we categorize visitors into four income groups based on their annual household income: $<\$60k$, $\$60$–$100k$, $\$100$–$150k$, and $>\$150k$. For each POI, we identify the income group that constitutes the largest proportion of visitors, thereby capturing the primary demographic profile of its customer base. We then employ the following model specification:
\begin{equation}Y_{it} = D_i \times  T_t \times \sum_{g=1}^{4} \beta_g \text{Income}_{igt} + u_i + \omega_{\text{year}} + \epsilon_{it}\end{equation} where $\text{Income}_{igt}$ is a binary variable that equals 1 if the majority of visitors to POI \textit{i} during time \textit{t} belong to most frequent income group \textit{g}, and 0 otherwise. This most frequent income group specification thus examines the heterogeneous treatment effects by visitor income.

\subsection*{Placebo test and staggered DID robustness} 
To test the hypothesis that differences in customer foot traffic to businesses are determined by the introduction of an EVCS rather than the impacts of spurious trends, we conduct a placebo test for the staggered adoption approach. In this placebo model, we randomly shuffle the opening dates of EVCSs, assigning each station a false installation date using the "sklearn.utils.shuffle" function in Scikit-learn. This placebo model reflects the idea that by assigning a false EVCS opening date, it is possible to assess whether the estimated treatment effect in the actual treatment group is biased. Using this random assignment process, ten placebo datasets were generated. For each of these placebo datasets, we apply the same staggered adoption approach to select control POIs and then perform the staggered DID analysis. We also conducted additional robustness checks to address concerns of self-selection within the staggered adoption strategy, including robustness to site selection based on nearest establishment, exploring more diverse pre-business size categories, wider treatment radius, and only considering visits made during the weekends. In this context, we repeat the main analysis, reflected by Equation (1), but using a larger, expanded sample of establishments, implementing the staggered adoption matching procedure.

\bibliography{sample}

\section*{Acknowledgements}

% We thank Placer.ai, who kindly provided us with the mobility data set for this research. 
T.Y. acknowledges support by the National Science Foundation under grant number CMMI 2425021. M.M.D.S. acknowledges support by the Japan Society for the Promotion of Science (JSPS) Grant-in-Aid for JSPS Fellows [grant number 24KJ1180]. The funders had no role in study design, data collection and analysis, decision to publish or preparation of the manuscript.

\section*{Author contributions statement}

M.M.D.S. and T.Y designed the algorithms, performed the analysis, and developed models. C.C performed part of the analysis and partially developed models. T.Y. and N.T. supervised the research. All authors wrote the paper. The company data were processed by T.Y. All authors had access to aggregated (non-individual) processed data. All authors reviewed the manuscript. 

\section*{Data availability}
The data that support the findings of this study are available from Placer.ai data platform, but restrictions apply to the availability of these data, which were used under the license for the current study and are therefore not publicly available. Information about how to request access to the data and its conditions and limitations can be found in \href{https://docs.placer.ai/reference/welcome-to-papi}{https://docs.placer.ai/reference/welcome-to-papi}. Data access requests should be submitted through Placer.ai's marketplace customer page \href{https://marketplace.placer.ai/}{https://marketplace.placer.ai/}, where the Sales team at Placer.ai may be contacted. Other data including the American Community Survey is available for download at \href{https://data.census.gov/}{https://data.census.gov/}, and Tiger shapefiles can be downloaded from the US Census Bureau \href{https://www.census.gov/programs-surveys/geography/guidance/tiger-data-products-guide.html/}{https://www.census.gov/programs-surveys/geography/guidance/tiger-data-products-guide.html}.

\section*{Code availability}
 The analysis was conducted using Python. Code to reproduce the main results in the figures from the aggregated data is publicly
 available on GitHub \href{https://github.com/Mavinds/Staggeredadoptionapproach}{https://github.com/Mavinds/Staggeredadoptionapproach}.

\section*{Competing interests statement}

The authors declare no competing interests.

\end{document}

\begin{center}
    {\large \textbf{Supplementary Information for}}\\[1em]
    {\LARGE \textbf{Causal spillover effects of electric vehicle charging station placement on local businesses: a staggered adoption study}}\\[2em]
    
    M. Mavin De Silva$^{1,2,3}$, Callie Clark$^{1,4}$, Tadachika Nakayama$^{2}$, Takahiro Yabe$^{1,5,*}$\\[1em]
    
    {\small
    $^1$Center for Urban Science and Progress (CUSP), Tandon School of Engineering, New York University, Brooklyn, NY, USA\\
    $^2$Extreme Energy-Density Research Institute, Nagaoka University of Technology, Nagaoka, Niigata 940-2188, Japan\\
    $^3$Department of Transport Management \& Logistics Engineering, Faculty of Engineering, University of Moratuwa, Katubedda 10400, Sri Lanka\\
    $^4$Marron Institute of Urban Management, New York University, Brooklyn, NY 11201, USA\\
    $^5$Department of Technology Management and Innovation, Tandon School of Engineering, New York University, Brooklyn, NY, USA\\[1em]
    *Corresponding author: \texttt{takahiroyabe@nyu.edu}}
\end{center}

\vspace{2em}

\renewcommand{\contentsname}{}

{\large\bfseries Supplementary Notes}

\tableofcontents

\newpage

\renewcommand{\thefigure}{S\arabic{figure}}
\renewcommand{\figurename}{Figure}

\renewcommand{\thetable}{S\arabic{table}}
\renewcommand{\tablename}{Table}

\renewcommand{\listfigurename}{}
{\large\bfseries List of Figures}
\listoffigures
\vspace{2em}

\renewcommand{\listtablename}{}
{\large\bfseries List of Tables}
\listoftables
\clearpage

\section{Data}
Our primary datasets include information on public charging station locations from the Department of Energy’s Alternative Fuels Data Center (AFDC), which is publicly available and covers over 65,000 stations as of January 2024\cite{USDOE2023}. Sociodemographic data were collected from the American Community Survey 2022 5-year estimates, detailing income, race/ethnicity, population statistics, and employment status for over 329 million Americans at the Census Block Group (CBG) level\cite{USCensus2022}. Additionally, data on auto-oriented intersections per square mile, total activity density, total road network density, and National Walkability Index score at the block group level were sourced from the Environmental Protection Agency’s (EPA) Smart Location Database\cite{USEPA2022}. Furthermore, NYC wide activity locations and mobility patterns were sourced from Placer.ai’s Global POIs Metadata \cite{PlacerAI2024} and the Placer API\cite{PlacerAPI2024b}, which is part of a Placer.ai Data Marketplace subscription. These datasets encompass over 10 million locations as of January 2024, classified under more than 600 NAICS codes and 250 major categories.

\subsection{Distribution of EV Charging Locations}
Five boroughs in New York City (NYC) are defined according to the NYC Government, with boundaries collected from the 2021 Census TIGER Shapefiles. \textbf{Figure S1} displays the selected NYC areas by presenting the number of EVCSs in each borough and \textbf{Figure S2} presents the number of EVCSs in California Bay Area (selected counties) over the same period.

\subsection{American Community Survey and Smart Location Detailed Tables}

The socioeconomic and demographic data on the census block groups were obtained from the American Community Survey 2022 five-year estimates\cite{USCensus2022}. Additionally, data on auto-oriented intersections per square mile, total activity density, and total road network density at the block group level (CBG), National Walkability Index scor were sourced from the Environmental Protection Agency’s (EPA) Smart Location Database. \textbf{Table S1} shows the detailed tables and features collected from each dataset.

\subsection{Selection of POI Categories} 
We applied multiple filtering steps to the Placer.ai Points of Interest (POIs) dataset to ensure relevance to charging station activity. Initially, we excluded seven high-level business categories defined by the North American Industry Classification System (NAICS) \cite{USCensus2022}, including ‘Administrative and Support and Waste Management and Remediation Services’, ‘Agriculture, Forestry, Fishing and Hunting’, ‘Mining, Quarrying, and Oil and Gas Extraction’, 'Gas Station / Garage', ‘Management of Companies and Enterprises’, ‘Utilities’, ‘Construction’, and ‘Manufacturing’. These categories collectively accounted for 2.5\% of the total POIs. To avoid redundancy, we also filtered out alternative fuel stations, including ‘EV Charging Stations’ and brands such as ‘ChargePoint Network Charging Station’, and ‘Tesla Destination Charger’. The selected POIs were then classified into elevan groups based on their NAICS high-level business categories, as follows: 

\begin{itemize}
    \item Apparel (e.g., Baby Store, Boutique, Bridal Shop, Clothing Store, Kids Store, Lingerie Store, Men's Store, Outlet Store, Shoe Store, Thrift / Vintage Store, Women's Store, Department Store, Accessories Store, Batik Shop, Jewelry Store, Leather Goods Store, Perfume Shop, Bike Shop, Fireworks Store, Fishing Store, Gun Shop, Hunting Supply, Outdoor Supply Store, Ski Shop, and Sporting Goods Shop)
    
    \item Beauty and Spa (e.g., Bath House, Cosmetics Shop, Health and Beauty Service, Massage Studio, Nail Salon, Piercing Parlor, Salon / Barbershop, Spa / Massage, Tanning Salon, and Tattoo Parlor)
    
    \item Buildings and Entertainment Centers (e.g., Buildings, Parks, Conference, Convention / Exhibition Centers, Rest Area, Dog Run, Entertainment Service, Meeting Room, Stoop Sale, CMBS – Commercial mortgage-backed securities)

    \item Dining (e.g., Breakfast, Coffee, Bakeries and Dessert Shops, Fast Food and QSR, Food Hall, and Restaurants)

    \item Groceries (e.g., Beer Store, Butcher, Candy Store, Cheese Shop, Chocolate Shop, Farmers Market, Fish Market, Food and Drink Shop, Food Service, Fruit and Vegetable Store, Gourmet Shop, Grocery Store, Health Food Store, Herbs and Spices Store, Liquor Store, Market, Organic Grocery, Sausage Shop, Supermarket, and Wine Shop)

    \item Home Improvement and Furnishings (e.g., Antique Shop, Carpet Store, Frame Store, Furniture / Home Store, Mattress Store, Construction and Landscaping, Garden Center, Hardware Store, Home Service, and Lighting Store)

    \item Hotels and Casinos (e.g., Agriturismo, Bed and Breakfast, Boarding House, Hostel, Motel, Mountain Hut, Ski Chalet, Ski Lodge, Vacation Rental, Casino, Hotel, Hotel Pool, Resort, and Ski Area)

    \item Leisure (e.g., Attractions, Bars and Pubs, Entertainment – Adult, Halls and Auditoriums, Nature and Landmarks, Outdoors and Recreation, Stadiums, Arenas and Athletic Fields, Theaters and Music Venue)

    \item Medical and Health (e.g., Alternative Medicine, Doctor and Health Professional Offices, Drugstores and Pharmacies, FSER – Free Standing Emergency Rooms, General Medical Centers and Hospitals, Home Health Care and Supply, Medical and Recreational Cannabis Dispensaries, Medical Centers and Services, Medical Laboratories, Specialty Hospitals, Urgent Care, Vocational Rehabilitation Services, and General Social Services)

    \item Shopping Centers (e.g., Paper / Office Supplies Store, Stationery Store, Community Shopping Centers, Shopping Center, Factory Outlets, Outlet Mall, Lifestyle Centers, Neighborhood Centers, Power Centers, Mall, Regional Malls, Strip / Convenience, Super-Regional Malls, Big Box Store, Warehouse Store, and Warehouse Club)

    \item Shops and Services (e.g., Banks and Financial Services, Car Shops and Services, Car Wash Services, Discount and Dollar Stores, Convenience Stores, Hobbies, Gifts and Crafts, Pet Stores and Services, Shipping and Storage, and Other Retail Stores and Services)
\end{itemize}

To analyze the association between EVCSs and activity places, we identified those eleven specific categories of POIs using the “top category” and “sub-category” attributes from the Placer.ai POI dataset. The selection process followed a stepwise approach to ensure the inclusion of relevant and meaningful categories as described above. \textbf{Table S2} displays the number of remaining POIs across five boroughs in NYC and \textbf{Table S3} shows the POI distribution across California Bay Area. 

\section{POI Matching Strategies}
This section provides an overview of the matched data using descriptive statistics and maps to illustrate the three matching approaches: the staggered adoption approach, the non-staggered matching approach and the non-staggered PSM approach. Given that EVCS providers strategically deploy charging stations in locations with high consumer traffic, economic activity, or accessibility advantages, POIs near EVCS installations may inherently differ from those without nearby chargers \cite{Zheng2024,Babar2024}. In order to evaluate whether ignoring the intrinsic characteristics of EVCS site selection introduces bias into the causal inference, we compared the staggered adoption approach with a non-staggered matching approach and a non-staggered propensity score matching (PSM) approach in two different urban settings: NYC, characterized by high EVCS density, and California Bay Area, which has moderate urban density and high EV adoption rates. The staggered adoption approach accounts for the timing of EVCS installations, ensuring that control POIs resemble treatment POIs before the intervention. In the two non-staggered matching approaches, untreated POIs were designated as those located outside the 400m radius of any EVCS opened during the same period. In addition, non-staggered matching approaches pairs POIs based on key characteristics, such as location attributes and economic activity, to create comparable groups. The following descriptive statistics and visualizations highlight the effectiveness of these matching strategies in reducing differences between treatment and control groups, enabling a more accurate assessment of the impact of EVCS installations, particularly as EVCS deployment expands into new areas.

\subsection{Staggered Adoption Matching Approach}

We define the treatment unit as any POI that lies within a 400-meter radius of a newly installed EVCS. For each installation event, we define a treatment window centered around the installation date and restrict inclusion to POIs within this spatial radius. To avoid overlapping treatment events and preserve temporal independence, our function filters potential matches based on three main rules: First, we ensure that the POIs used as controls are not simultaneously or imminently exposed to future treatment, thereby maintaining the integrity of the control group. To ensure similar local economic conditions, we restrict matching to POIs located within the same borough \textit{b} and belonging to the same business category \textit{c}, such that $i, j \in \{\text{POIs}(b, c)\}$; Second, the control POI must be associated with a different EVCS than the one linked to the treated POI. This spatial separation minimizes contamination from overlapping or nearby treatment zones. Third, the control POI’s station must have been installed at least three months after the installation date of the treated POI’s station. This temporal gap ensures that control POIs are not already influenced by treatment effects during the matching period. The output is a set of potential control POIs that satisfy all of these criteria, along with their corresponding metadata. 

To ensure the validity of the Difference-in-Differences (DID) framework, a critical step involves verifying that treated and control POIs exhibit parallel trends in visitation prior to the treatment event\cite{Ito2024}. This is essential for isolating the causal impact of the intervention, in this case, the installation of an EVCS. For each treated POI, we examine visitation patterns across three distinct time windows relative to the installation date: the early pre-treatment period (6 to 3 months prior), the immediate pre-treatment period (the 3 months leading up to the installation), and the post-treatment period (the 3 months following installation). We calculate the average visit counts during the early and immediate pre-treatment periods and compute their difference, which serves as a proxy for the pre-treatment visitation trend of the treated POI. This same calculation is then performed for each potential control POI, using the treated POI's installation date as the reference point to maintain temporal consistency. By comparing these differences, we assess how closely the control POIs mirror the untreated visitation trajectory of the treated POI. 

In order to match each treated POI with the most comparable control POI based on their pre-treatment visitation trends, we implement a greedy nearest neighbor algorithm that selects control POIs with visitation trends most similar to those of the treated POIs in the period prior to EVCS installation\cite{Babar2024}. Specifically, for each treated POI $i$, we compute the absolute difference in pre-treatment trends, measured as the change in average visits between the early pre-treatment period (6 to 3 months prior) and the immediate pre-treatment period (3 months prior to installation), relative to every candidate control POI $j$. The control POI that minimizes this absolute difference is selected as the match: $\arg\min_j \left| \text{DID}_i - \text{DID}_j \right|$. This procedure is implemented using the \texttt{idxmin()} function from Python’s Pandas library, which efficiently identifies the index of the minimum value. The output is a matched dataset in which each treated POI is paired with a control POI exhibiting a closely aligned pre-treatment trend. Notably, the greedy nature of the algorithm prevents the reuse of control POIs unless explicitly allowed, reducing dependency among matched pairs and avoiding over representation of certain control units.

\textbf{Figure S3} and \textbf{Figure S4} visualize the structural properties of matched POI pairs, categorized by treatment type and distance proximity, using the staggered adoption approach in NYC and California Bay Area, respectively. Yellow points represent treated POIs, while green points indicate control POIs, and blue points denote POIs that were initially treated but later served as controls in subsequent time periods. This visualization helps illustrate the spatial distribution and selection process within the staggered matching approach, highlighting how control and treatment groups are constructed over time. \textbf{Table S4} and \textbf{Table S5} provide descriptive statistics for the data obtained using the staggered adoption matching approach in experimental settings.  

\subsection{Non-staggered Matching Approach}
In the non-staggered matching approach, we adopt the same multi-step matching procedure used in the staggered setting, with one key distinction: untreated POIs are defined as those located outside the 400-meter radius of any EVCS that opened during the same period. This definition ensures that control POIs are spatially insulated from the treatment event, allowing for a cleaner comparison. As in the staggered case, we begin by defining the treatment unit as any POI located within a 400-meter radius of an EVCS installation. For each treated POI, we construct a treatment window centered around the installation date and identify all POIs outside the treatment radius as candidate controls. To ensure contextual similarity and limit unobserved heterogeneity, we restrict matches to POIs located within the same borough \textit{b} and belonging to the same business category \textit{c}, such that $i, j \in \{\text{POIs}(b, c)\}$. We then validate that treated and control POIs exhibit parallel visitation trends before the treatment, which is critical to the validity of the DID framework. Specifically, we calculate visit averages over three time windows relative to the EVCS installation: an early pre-treatment period (6 to 3 months prior), an immediate pre-treatment period (the 3 months before installation), and a post-treatment period (the 3 months following installation). The difference in average visits between the early and immediate pre-treatment periods serves as a proxy for the POI’s pre-treatment trend. This same calculation is applied to all control POIs using the treated POI’s installation date as the anchor.

 To perform the final match, we implement a greedy nearest neighbor algorithm to pair each treated POI with the control POI whose pre-treatment trend most closely matches its own\cite{Babar2024}. For each treated POI  $i$, we compute the absolute difference in visitation trends relative to all potential control POIs  $j$, and select the match that minimizes this difference: $\arg\min_j \left| \text{DID}_i - \text{DID}_j \right|$. This algorithm is implemented using the \texttt{idxmin()} function from Python’s Pandas library to efficiently identify the best match. \textbf{Figure S5} and \textbf{Figure S6}  represent the structural properties of matched POI pairs, categorized by treatment type and distance proximity, using the non-staggered matching approach. Here, yellow points represent treated POIs, while green points indicate control POIs. \textbf{Table S6} and \textbf{Table S7} provide descriptive statistics for the data obtained using the non-staggered matching approach. 

\subsection{Non-staggered Propensity Score Matching}
In the non-staggered PSM approach, we aimed to construct a comparison group of control POIs that closely resembled the treated POIs in observable characteristics but were not exposed to treatment\cite{Zheng2024}. Similar to the exact-nearest matching approach, treatment units were defined as POIs located within a 400m radius of an EVCS installed during the study period. Crucially, control units were defined as POIs that remained outside the 400m radius of any EVCS installed during the same period, ensuring that they remained unaffected by the treatment event and thus suitable for comparison. To account for selection bias and improve the balance of observed covariates between treated and untreated units, we employed a PSM framework that pairs treated POIs with control POIs exhibiting similar characteristics. These characteristics included: (1) Built environment factors, such as population density, the ratio of auto-oriented road length to total road network length, and walkability, which influence accessibility and activity levels—key considerations for EVCS siting; (2) Socio-demographic variables, including median income, unemployment rate, and racial composition, which capture neighborhood-level economic conditions and EV adoption potential; and (3) POI-level variables, such as the average number of customer visits prior to the installation date, which reflect underlying consumer activity and commercial attractiveness. 

The matching process was conducted in two steps. First, we performed exact matching based on same borough \textit{b} and POI category \textit{c}, ensuring that treated and control POIs operated within the same sector and geographic region. This step helped control for both industry-specific trends and local economic or policy conditions that might influence visitation patterns and EVCS siting decisions. Second, we implemented nearest-neighbor propensity score matching, using a logistic regression model to estimate each POI's probability of being treated based on the aforementioned covariates. The model was fitted using the statsmodels.formula.api.logit function in Python. The resulting propensity scores were then used in a nearest-neighbor algorithm implemented via Scikit-learn’s sklearn.neighbors.KNeighborsRegressor, to identify control POIs whose likelihood of treatment closely matched that of treated POIs. \textbf{Figure S7} and \textbf{Figure S8} show the structural properties of matched POI pairs, categorized by treatment type and distance proximity, using the non-staggered PSM matching approach. Here, yellow points represent treated POIs, while green points indicate control POIs. \textbf{Table S8} and \textbf{Table S9} provide descriptive statistics for the data obtained using the non-staggered PSM matching approach. 

\subsection{Robustness against Choice of Matching Approaches}
\textbf{Figure S9} and \textbf{Figure S10} illustrate the covaraite balance and mean distance between matched pairs under the staggered adoption approach (purple line), exact-nearest matching (orange line), and the non-staggered PSM approach (green line) for the five boroughs across NYC and California Bay Area, respectively. The difference in distance means is statistically significant $(\text{p-value} < 0.001)$, underscoring the effectiveness of using the staggered nature of EVCS installations to establish comparable groups, particularly in EVCS dense areas of NYC.

\section{Staggered Difference in Difference}
This section explores the key determinants influencing the staggered Difference-in-Differences (DID) estimates and examines the heterogeneity of treatment effects across different dimensions. The analysis aims to provide a deeper understanding of how the impact of EVCS installations varies depending on spatial proximity, POI characteristics, and socioeconomic factors. 

\subsection{Impact of EVCS installations on customer count}
In this sub-section, we compare the baseline staggered DID results with those obtained using the non-staggered matching approaches and placebo tests. The baseline model specification is defined as follows: \begin{equation}
Y_{it} = \beta D_i \times T_t + u_i + \omega_{\text{year}} + \epsilon_{it}\end{equation} In this equation, $Y_{it}$ represents the outcome variable, which is the number of customers at POI \textit{i} during time period \textit{t}. $\beta$ represents the treatment effect, quantifying the impact of adding a new EVCS on the number of customers at the respective POI.  The term $D_{i}$ is a binary treatment indicator that takes on the value of 1 for treated establishments when an EVCS opens for service nearby. Similarly, $T_{t}$ is a binary time indicator that shifts from zero to one in the post-treatment period. $u_i$ denotes the POI fixed effects. $\omega_{\text{year}}$ represents the year fixed effects. $\epsilon_{it}$ represents the error term. \textbf{Table S10} and \textbf{Table S11} assess the robustness of the baseline model findings across the two urban settings, respectively. The results of the placebo test serve as a validation check, helping to distinguish genuine treatment effects from spurious correlations.

\subsection{Heterogeneity of effects by POI categories}
In this sub-section, we explore the heterogeneity of the effects by POI categories, using the following model specification: 
\begin{equation}Y_{it} = D_i \times T_t \times \sum_{c=1}^{11} \beta_c \text{Cat}_{ict} + u_i + \omega_{\text{year}} + \epsilon_{it}\end{equation} Here, $\text{Cat}_{ict}$ represents customer visits for POI Category \textit{c} at POI \textit{i} during time period \textit{t}, where \textit{\textbf{c}} can take on one of eleven possibilities: Apparel, Beauty and Spa, Buildings and Entertainment Centers, Dining, Groceries, Home Improvements and Furnishings, Hotels and Casinos, Leisure, Medical and Health, Shopping Centers, and Shops and Services. $\beta_{c}$ thus capture the treatment effects associated with each POI type. \textbf{Table S12}  and \textbf{Table S13} represent the treatment effect specific to a particular type of POI while comparing the baseline model with the non-staggered matching approaches and placebo test results across the two urban settings, respectively.

\subsection{Variation in treatment effects by distance}
In this sub-section, by categorizing POIs into four distance bins spanning from 0 to 400m, with 100m increments, we assess whether the magnitude of treatment effects diminishes or intensifies with increasing distance from the EVCS as follows \textbf{Table S14} for NYC and \textbf{Table S15} across six counties in California: \begin{equation}Y_{it} = D_i \times  T_t \times \sum_{d=1}^{4} \beta_d \text{Dis}_{idt} + u_i + \omega_{\text{year}} + \epsilon_{it}\end{equation} where, $\text{Dis}_{idt}$ is a binary dummy variable that takes the value of 1 if an EVCS was opened during the study period time \textit{t} and falls within the distance bin d of POI \textit{i}.

\subsection{Effects on populations from different income groups}
In this sub-section, we explore the heterogeneity of the effects from difference income groups, by categorizing visitors into four income groups based on their annual household income: $<\$60k$, $\$60$–$100k$, $\$100$–$150k$, and $>\$150k$. For each POI, we identify the income group that constitutes the largest proportion of visitors, thereby capturing the primary demographic profile of its customer base. We then employ the following model specification:
\begin{equation}Y_{it} = D_i \times  T_t \times \sum_{g=1}^{4} \beta_g \text{Income}_{igt} + u_i + \omega_{\text{year}} + \epsilon_{it}\end{equation} where $\text{Income}_{igt}$ is a binary variable that equals 1 if the majority of visitors to POI \textit{i} during time \textit{t} belong to most frequent income group \textit{g}, and 0 otherwise. The estimated $\beta$ values of the heterogeneous EVCS effects from different income groups by comparing the three model results are presented in \textbf{Table S16} for NYC and \textbf{Table S17} across six counties in California, respectively.

\begin{figure}
    \centering
    \includegraphics[width=1\linewidth]{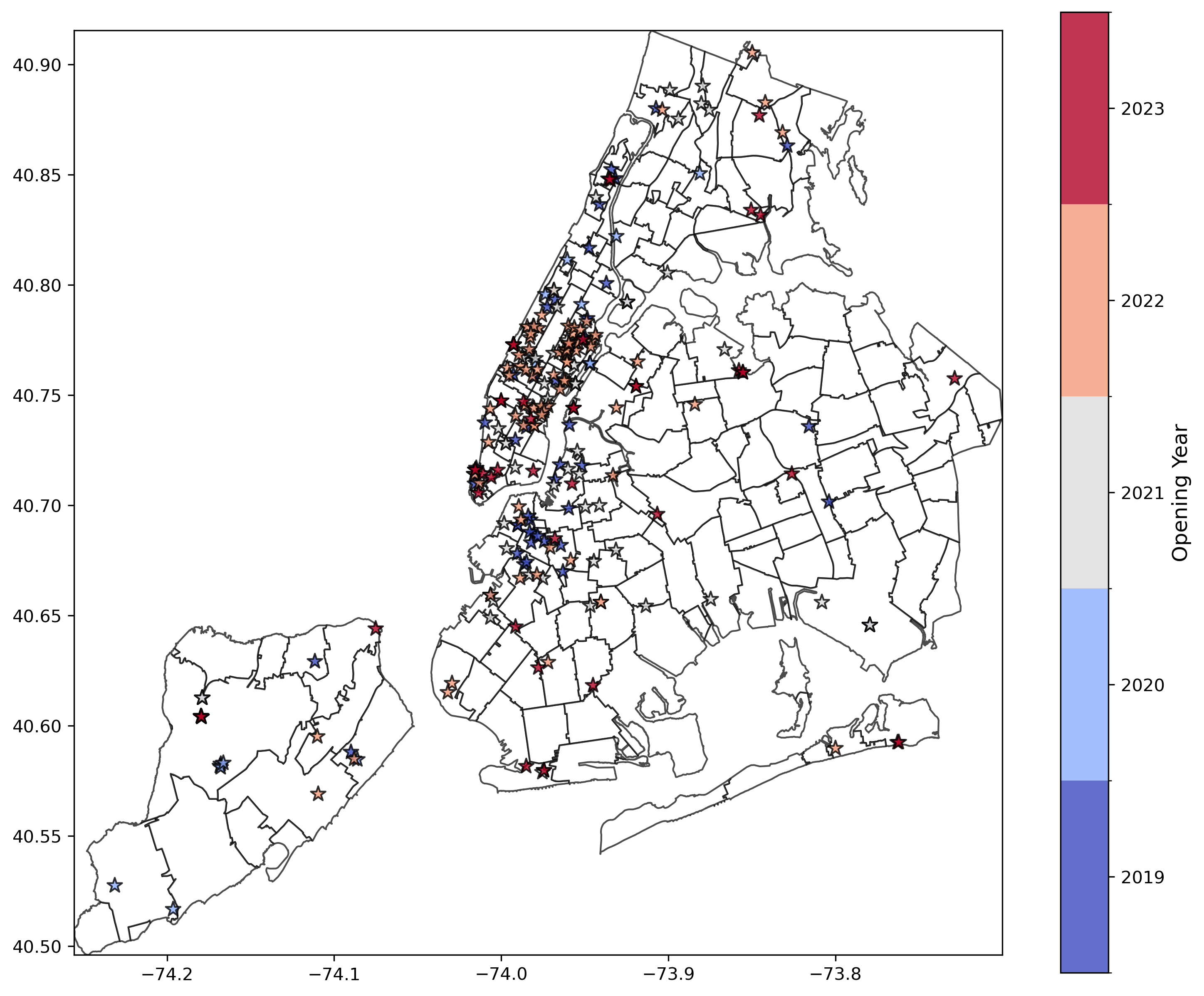}
    \caption{Locations of EVCS across five boroughs in NYC}
    \label{fig:enter-label1}
\end{figure}

\begin{figure}
    \centering
    \includegraphics[width=1\linewidth]{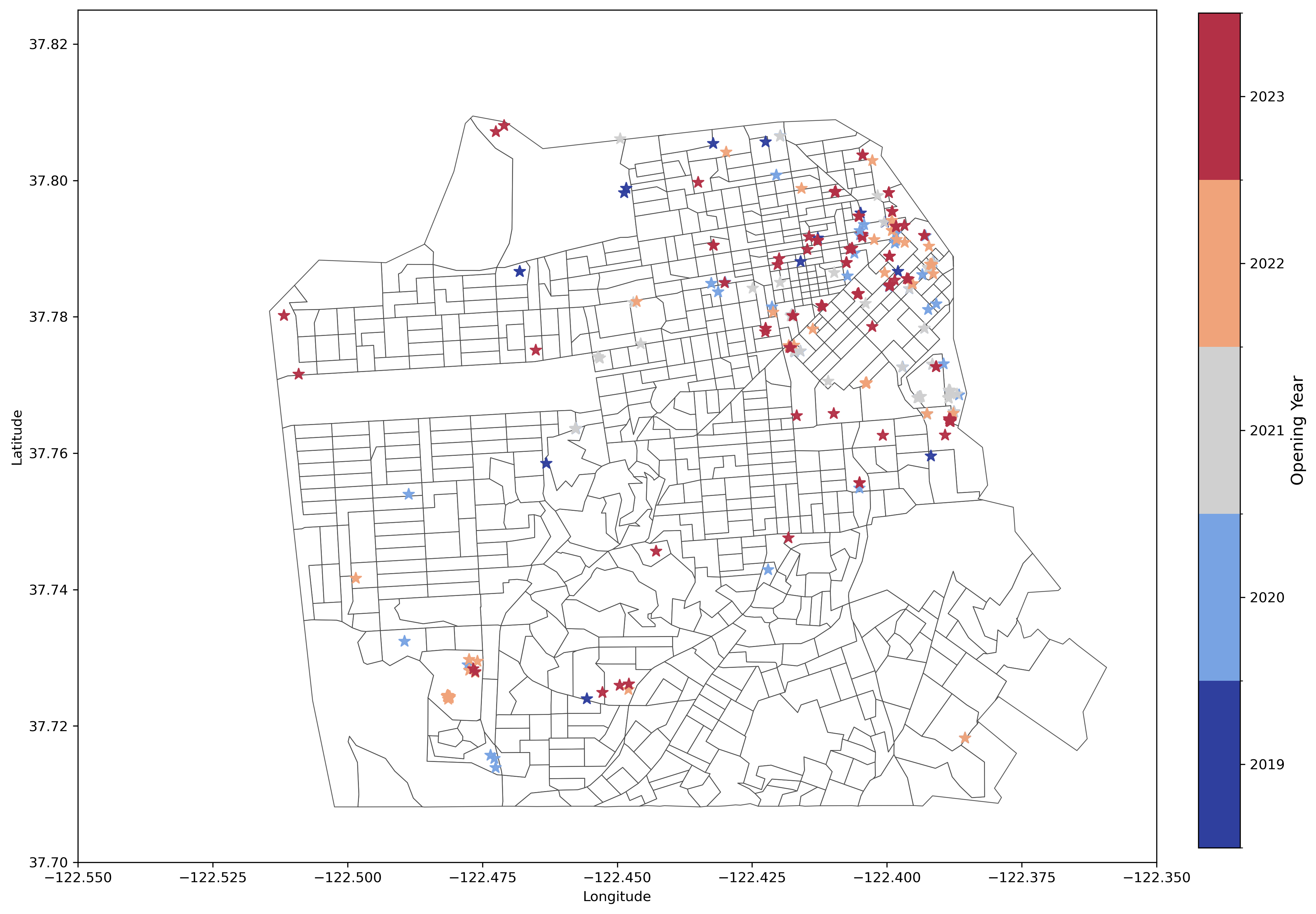}
    \caption{Locations of EVCS in San Francisco County}
    \label{fig:placeholder1}
\end{figure}

\begin{figure}
    \centering
    \includegraphics[width=1\linewidth]{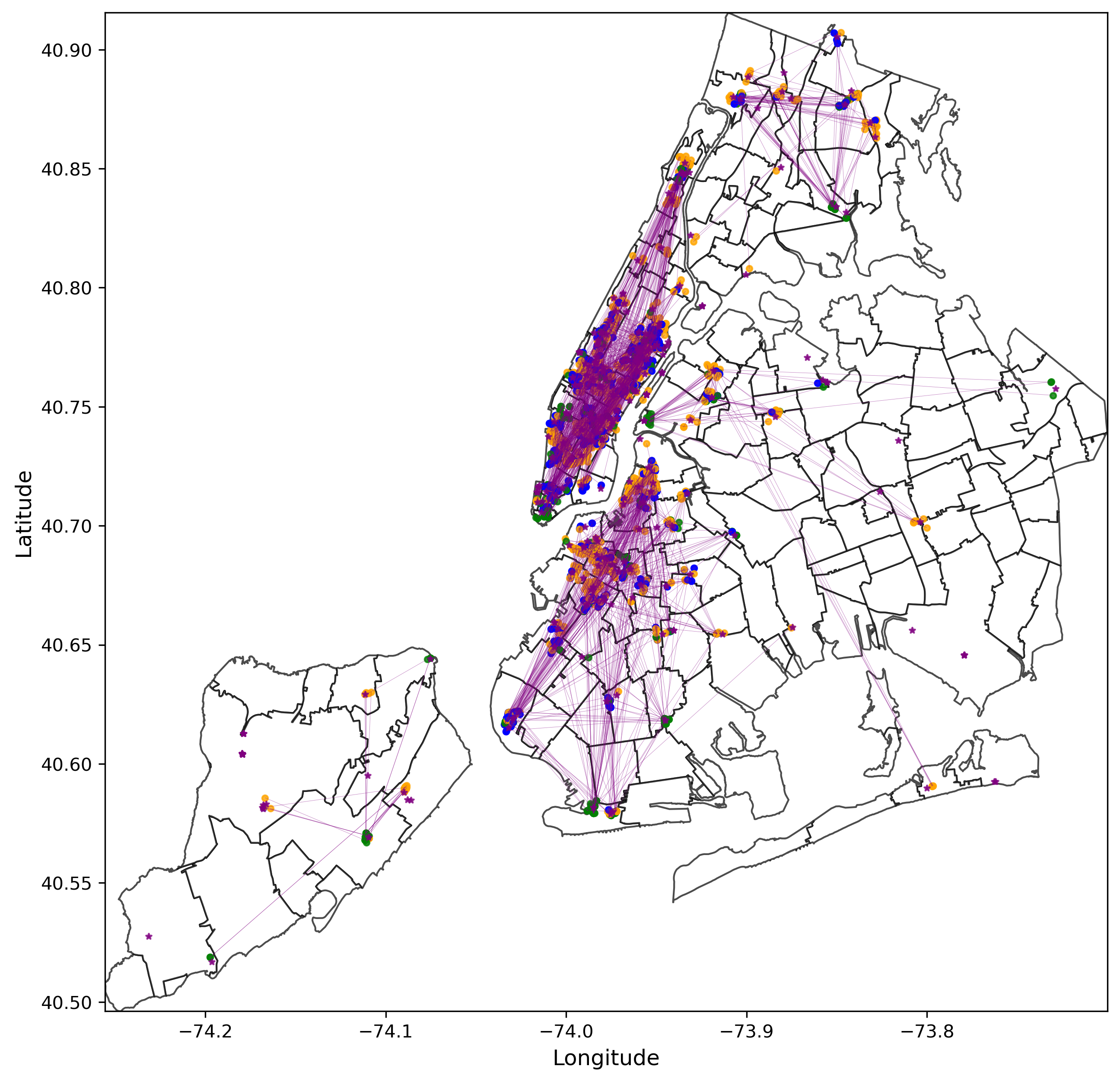}
    \caption{Spatial distribution of matched POI pairs across treatment types and distance proximity under the staggered adoption approach in NYC}
    \label{fig:enter-label2}
\end{figure}

\begin{figure}
    \centering
    \includegraphics[width=1\linewidth]{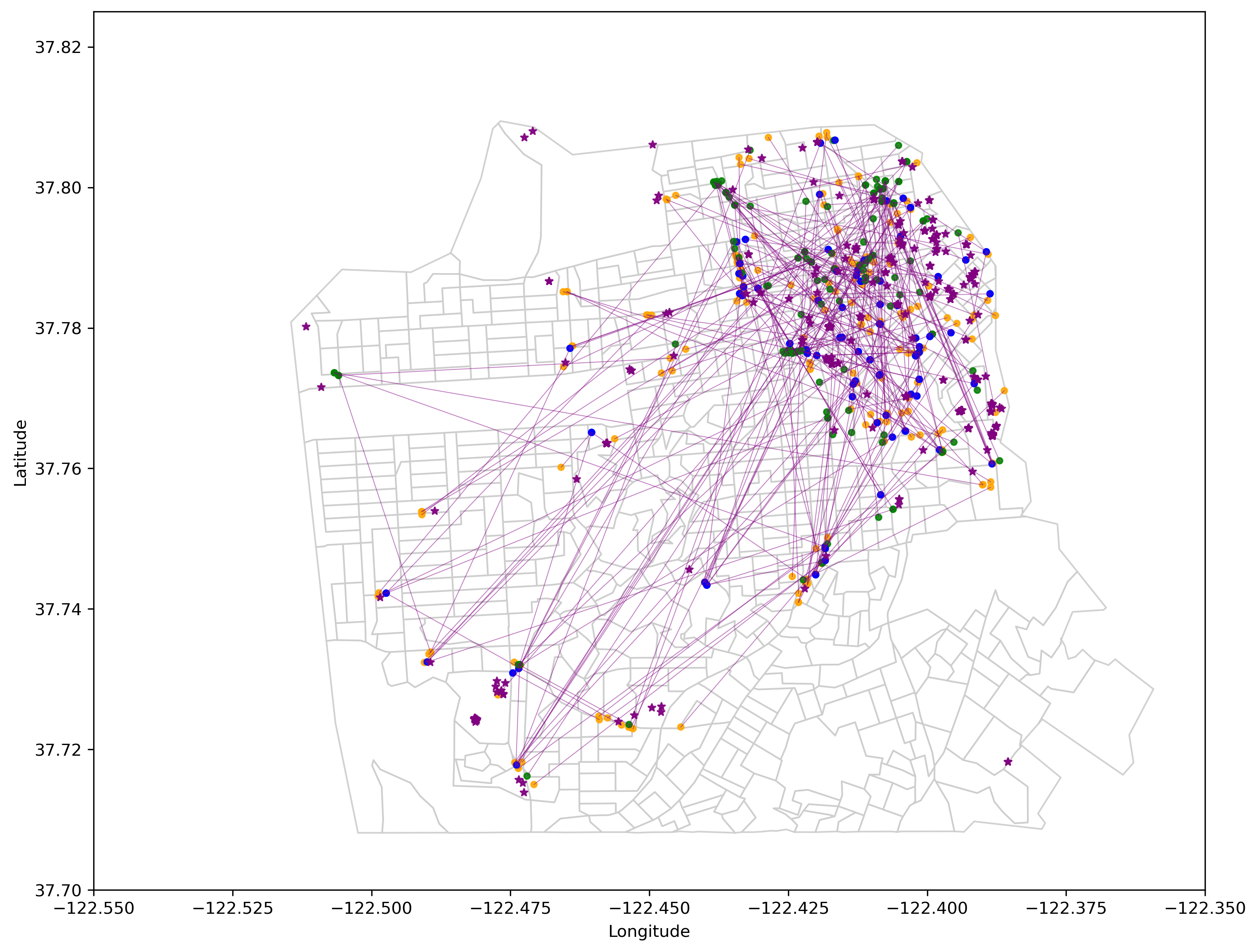}
    \caption{Spatial distribution of matched POI pairs across treatment types and distance proximity under the staggered adoption approach in San Francisco county}
    \label{fig:placeholder2}
\end{figure}

\begin{figure}
    \centering
    \includegraphics[width=1\linewidth]{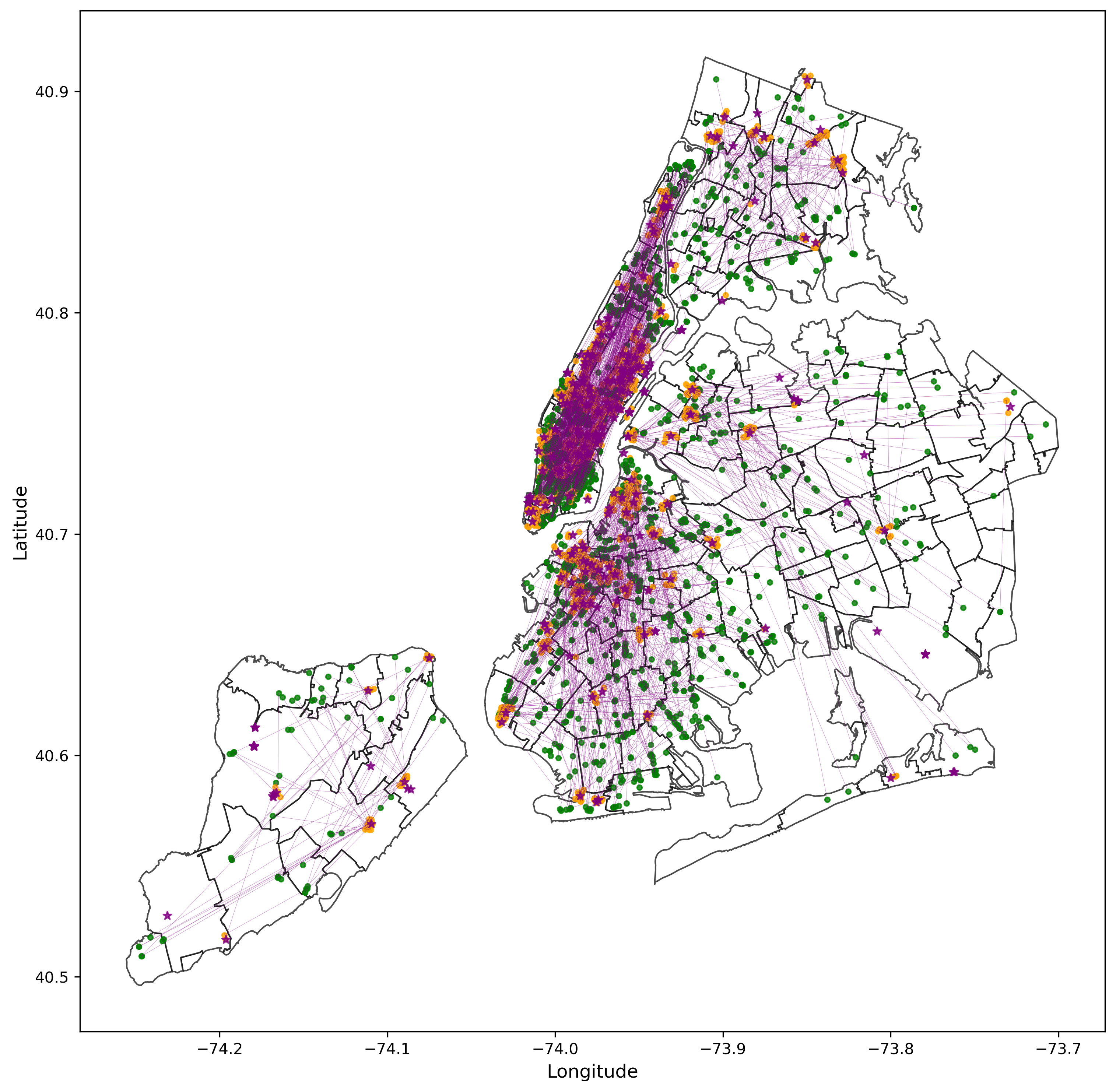}
    \caption{Spatial distribution of matched POI pairs across treatment types and distance proximity under the non-staggered matching approach in NYC}
    \label{fig:enter-label3}
\end{figure}

\begin{figure}
    \centering
    \includegraphics[width=1\linewidth]{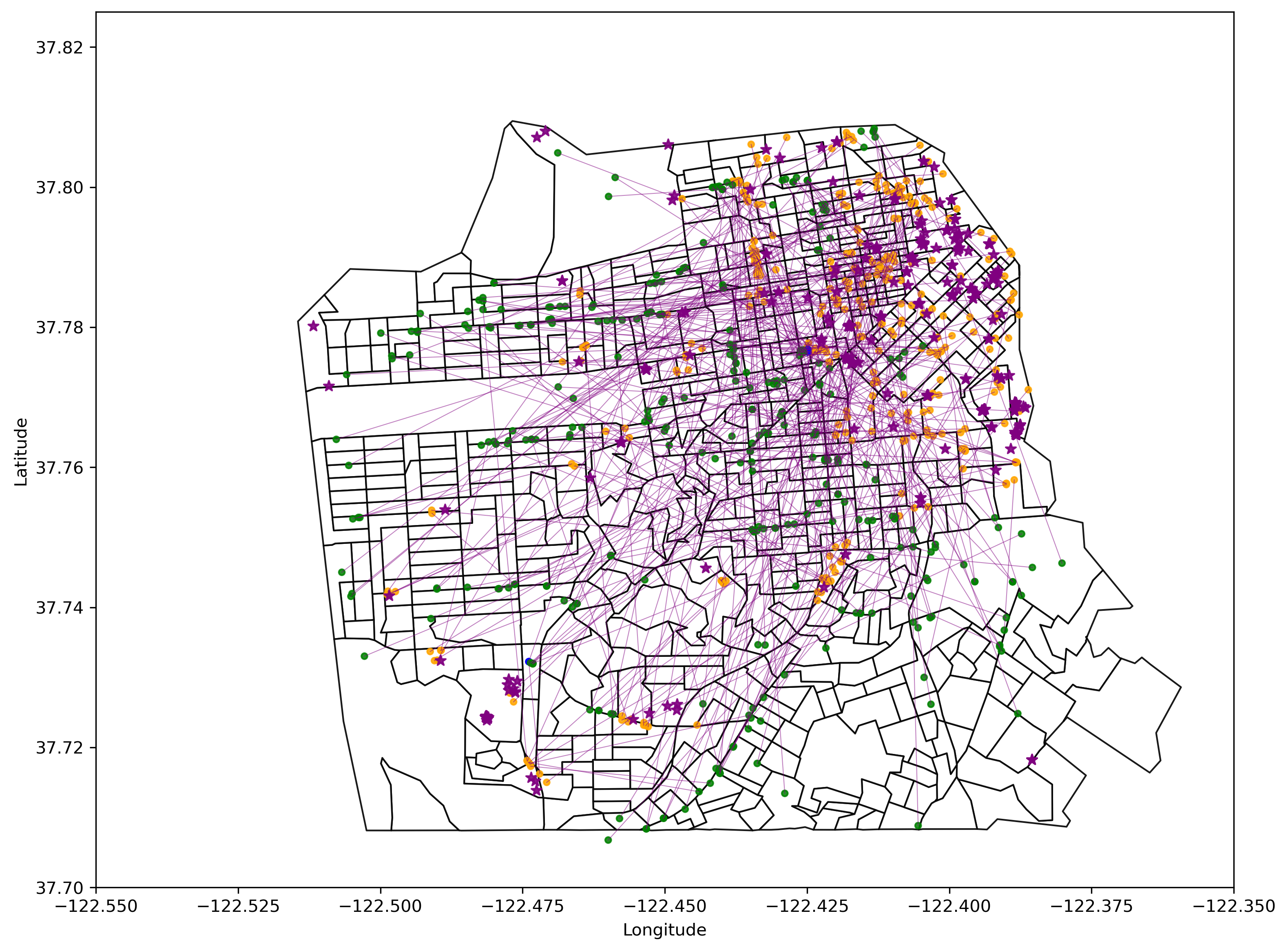}
    \caption{Spatial distribution of matched POI pairs across treatment types and distance proximity under the non-staggered matching approach in San Francisco county}
    \label{fig:placeholder3}
\end{figure}

\begin{figure}
    \centering
    \includegraphics[width=1\linewidth]{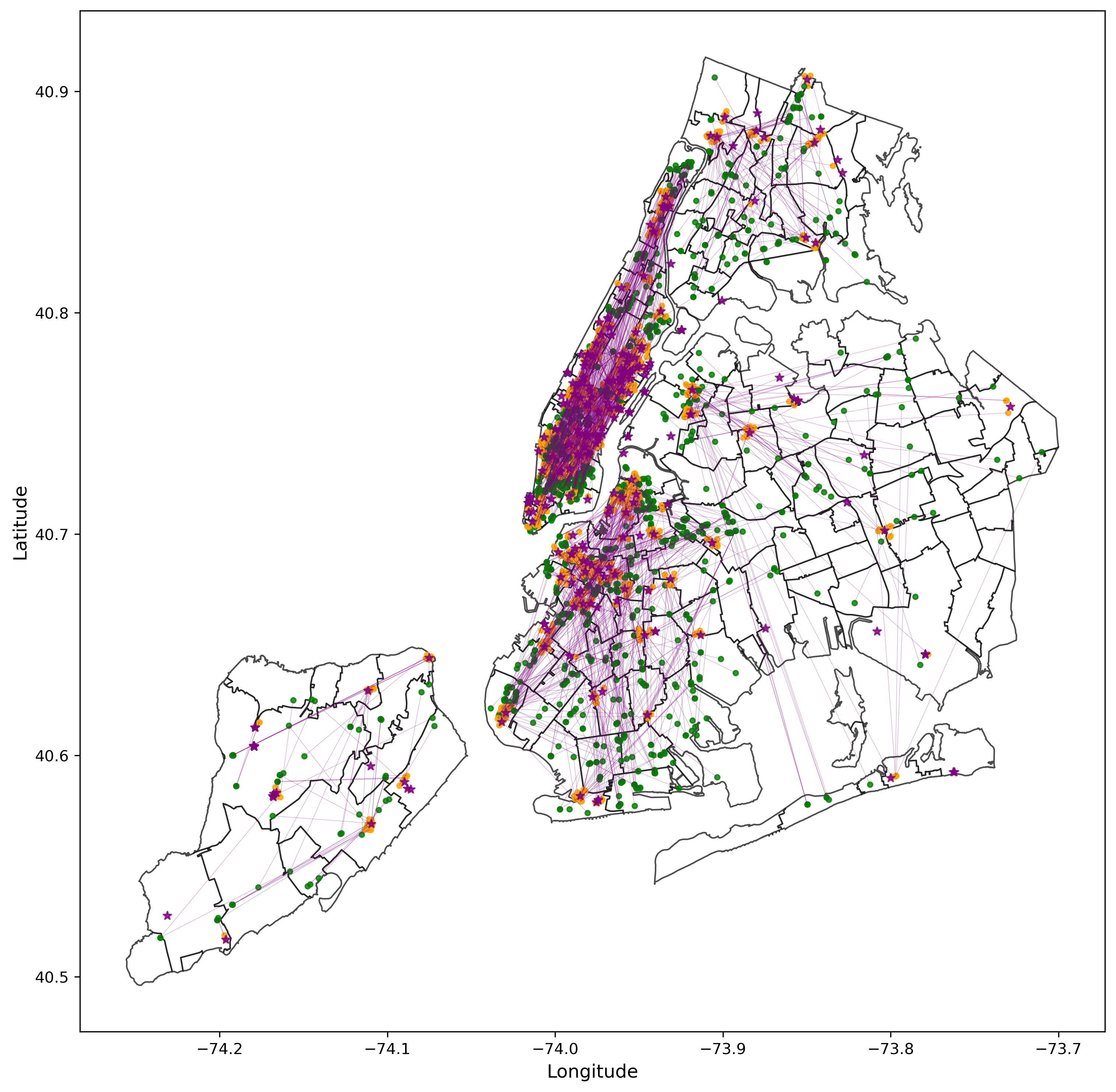}
    \caption{Spatial distribution of matched POI pairs across treatment types and distance proximity under the non-staggered PSM approach in NYC}
    \label{fig:enter-label4}
\end{figure}

\begin{figure}
    \centering
    \includegraphics[width=1\linewidth]{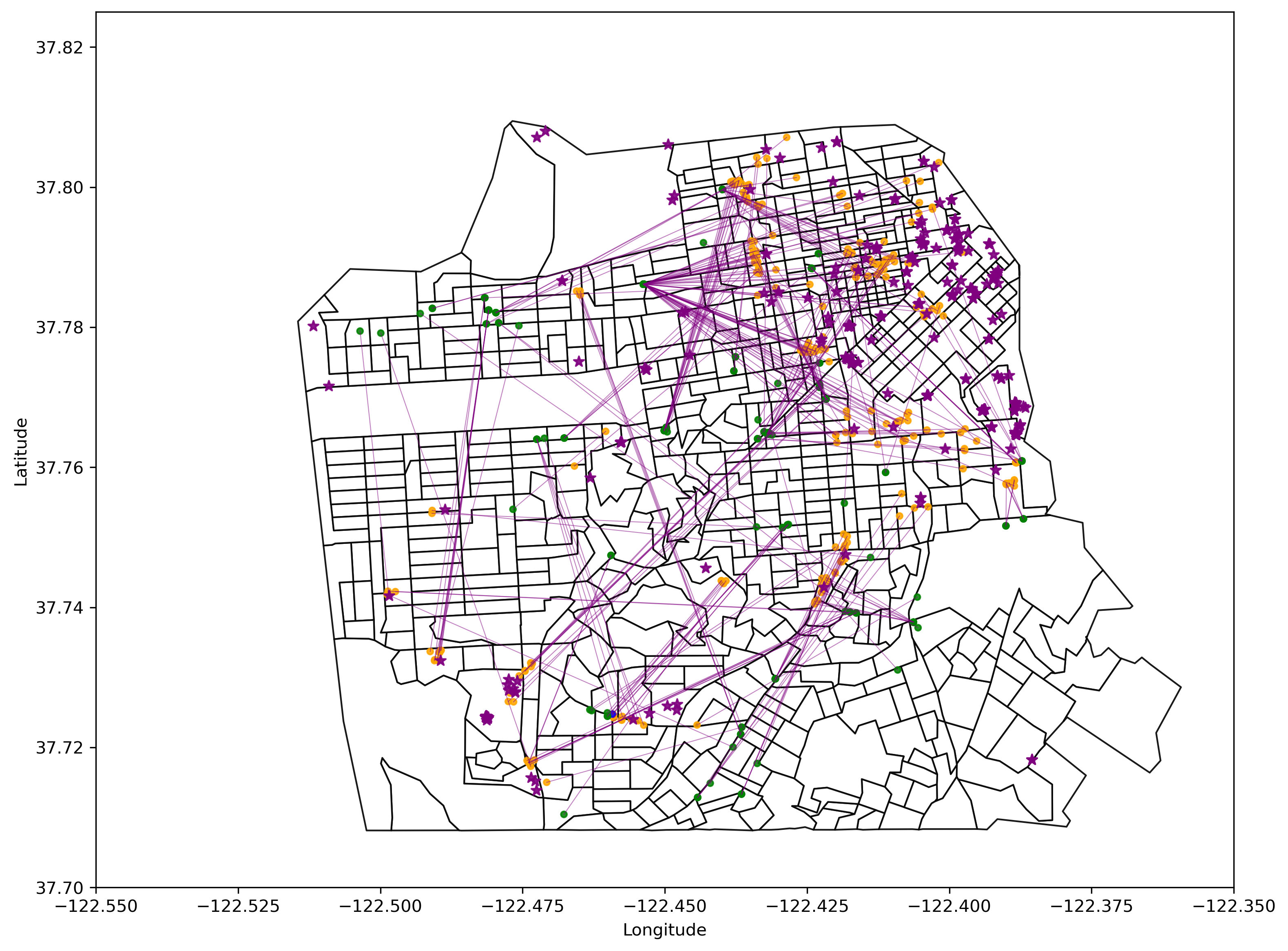}
    \caption{Spatial distribution of matched POI pairs across treatment types and distance proximity under the non-staggered PSM approach in San Francisco county}
    \label{fig:placeholder4}
\end{figure}

\begin{figure}
    \centering
    \includegraphics[width=1\linewidth]{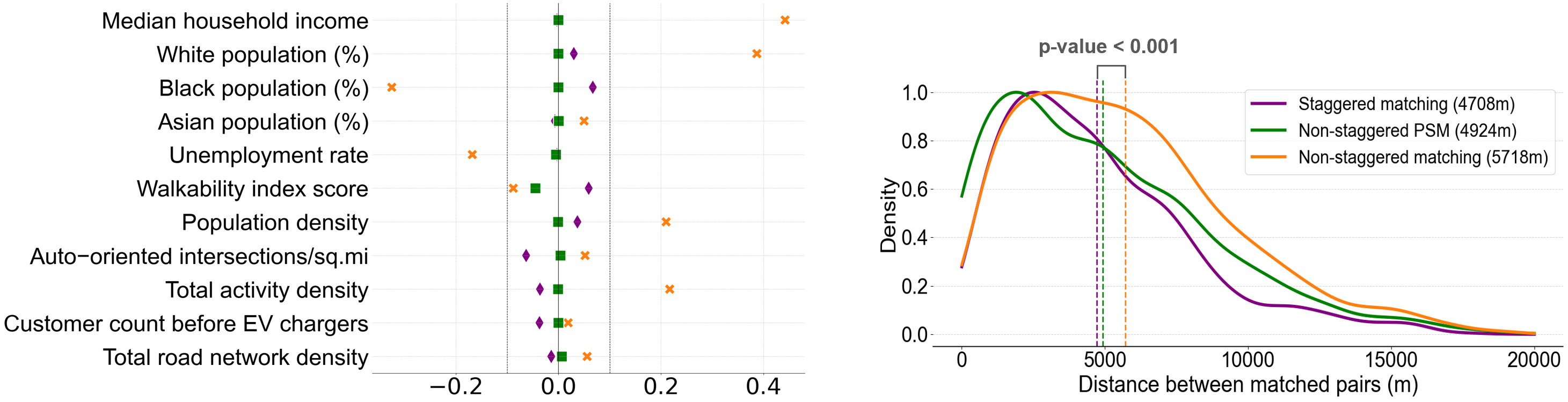}
    \caption{Mean distance and covaraite balance between matched pairs in each borough in NYC under the staggered adoption approach, non-staggered non-staggered matching approach and the non-staggered matching PSM approach}
    \label{fig:placeholder5}
\end{figure}

\begin{figure}
    \centering
    \includegraphics[width=1\linewidth]{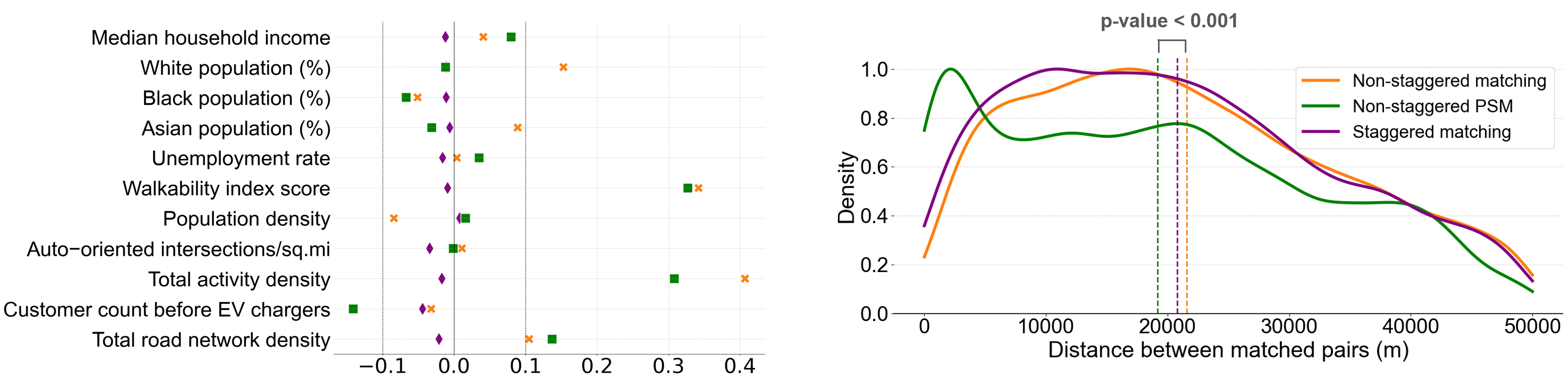}
    \caption{Mean distance and covaraite balance between matched pairs across California Bay Area under the staggered adoption approach, non-staggered non-staggered matching approach and the non-staggered matching PSM approach}
    \label{fig:placeholder6}
\end{figure}

\begin{table}[h]
\centering
\small
\caption{ Detailed Tables from American Community Survey (ACS) and Smart Location Database (SDL)}
\label{tab:S2}
\begin{tabular}{lcc}
\hline
\textbf{Dataset and Code} & \textbf{Feature Name} & \textbf{Table Description} \\
\multirow{3}{*}{ACS-B02001}  
  & White \%  & \multirow{3}{*}{\makecell{Percentage of individuals in a \\ census block group who list \\ their race as the given category}}  \\
  & Black \%  &  \\
  & Asian \%  &  \\
ACS-B19013  & Median Household Income & \makecell{Median household income  (\$) \\ (inflation-adjusted dollars)}  \\
ACS-B23025  & Unemployment Rate \% & \makecell{Percentage of individuals \\ who are unemployed} \\
SLD-NatWalkInd  &  Walkability Index Score & \makecell{Components of the built environment \\ that influence the likelihood of walking} \\
SLD-D1b  &  Population Density  & \makecell{Gross population density (people/acre) \\ on unprotected land } \\
SLD-D3bao  &   Auto-oriented Intersections /sqm & \makecell{Intersection density in terms of \\ auto-oriented intersections /sqm} \\
SLD-D1d  &  Total Activity Density  & \makecell{Gross activity density (employment + HUs) \\ on unprotected land } \\
SLD-D3a  &  Total Road Network Density   & \makecell{Total street network density \\ on total land } \\
\hline
\end{tabular}
\end{table}

\begin{table}
\centering
\caption{Number of places in the Placer.ai dataset across the five boroughs in NYC}
\label{tab:S3}
\begin{tabular}{lccccc}
\hline
\textbf{Place category} & \textbf{Bronx} & \textbf{Brooklyn} & \textbf{Manhattan} & \textbf{Queens} & \textbf{Staten Island} \\
\hline
Apparel & 12 & 42 & 62 & 5 & 12 \\
Beauty \& Spa & 11 & 30 & 77 & 11 & 2 \\
\multirow{2}{*}{Buildings} & \multirow{2}{*}{3} & \multirow{2}{*}{21} & \multirow{2}{*}{199} & \multirow{2}{*}{7} & \multirow{2}{*}{1} \\
\multirow{2}{*}{and Entertainment Centers} & & & & & \\
\vspace{-5pt} \\ % Adds space between multirows
Dining & 51 & 313 & 811 & 64 & 25 \\
Groceries & 20 & 74 & 106 & 14 & 1 \\
\multirow{2}{*}{Home Improvements} & \multirow{2}{*}{4} & \multirow{2}{*}{27} & \multirow{2}{*}{38} & \multirow{2}{*}{14} & \multirow{2}{*}{-} \\
\multirow{2}{*}{and Furnishings} & & & & & \\
\vspace{-5pt}\\
Hotels \& Casinos & 2 & 32 & 173 & 10 & 9 \\
Leisure & 5 & 112 & 332 & 20 & 3 \\
Medical \& Health & 10 & 62 & 93 & 4 & 8 \\
Shopping Centers & 9 & 31 & 97 & 6 & 8 \\
Shops \& Services & 55 & 178 & 235 & 50 & 19 \\
\hline
\end{tabular}
\end{table}

\begin{table}
\centering
\caption{Number of places in the Placer.ai dataset across California Bay Area}
\label{tab:S4}
\begin{tabularx}{\textwidth}{l *{6}{>{\centering\arraybackslash}X}}
\hline
\textbf{Place category} & \textbf{Alameda} & \textbf{LA} & \textbf{Orange} & \textbf{San Francisco} & \textbf{San Mateo} & \textbf{Santa Clara} \\
\hline
Apparel & 26 & 307 & 144 & 9 & 21 & 77 \\
Beauty \& Spa & 40 & 400 & 184 & 10 & 29 & 104 \\
Buildings & 4 & 51 & 9 & 6 & 9 & 6 \\
Entertainment Centers & 272 & 1744 & 653 & 89 & 179 & 404 \\
Dining & 22 & 220 & 50 & 10 & 20 & 24 \\
Groceries & 29 & 210 & 74 & 20 & 34 & 48 \\
Home Improvements & 16 & 80 & 41 & 18 & 17 & 26 \\
Furnishings & 34 & 225 & 60 & 20 & 17 & 35 \\
Hotels \& Casinos & 36 & 362 & 128 & 8 & 31 & 55 \\
Leisure & 5 & 72 & 24 & 3 & 2 & 8 \\
Medical \& Health & 273 & 1572 & 471 & 72 & 152 & 289 \\
\hline
\end{tabularx}
\end{table}

\begin{table}[h!]
    \centering
    \renewcommand{\arraystretch}{1.2}
    \caption{Descriptive statistics for staggered adoption matching pairs in NYC}
    \resizebox{\textwidth}{!}{%
    \begin{tabular}{lccccc}
        \toprule
        \textbf{Variable} & \textbf{N} & \textbf{Mean} & \textbf{Std. dev} & \textbf{Min} & \textbf{Max} \\
        \midrule
        \multicolumn{6}{l}{\textbf{Treated group}} \\
        Monthly number of customers per POI (in 1000s) & 1733 & 0.29 & 0.68 & 0.00 & 10.36 \\
        Median household income (in \$1000s) & 1733 & 142.67 & 60.45 & 11.83 & 250.00 \\
        Race: White population (\%) & 1733 & 63.06 & 23.47 & 0.00 & 100.00 \\
        Race: Black or African American population (\%) & 1733 & 8.97 & 17.64 & 0.00 & 100.00 \\
        Race: Asian population (\%) & 1733 & 11.11 & 10.31 & 0.00 & 73.80 \\
        Unemployment rate (\%) & 1733 & 5.82 & 6.45 & 0.00 & 53.85 \\
        Walkability index score & 1733 & 14.70 & 2.70 & 5.33 & 19.67 \\
        Population density (people/km$^2$) & 1733 & 123.56 & 81.62 & 0.00 & 566.00 \\
        Road density & 1733 & 3.03 & 6.18 & 0.00 & 78.52 \\
        Intersection density (intersections/km$^2$) & 1733 & 13.49 & 39.52 & 0.00 & 525.44 \\
        Activity density & 1733 & 235.41 & 299.39 & 0.00 & 2437.36 \\
        Road network & 1733 & 32.25 & 11.39 & 2.05 & 112.91 \\
        \midrule
        \multicolumn{6}{l}{\textbf{Matched control group}} \\
        Monthly number of customers per POI (in 1000s) & 912 & 0.32 & 0.75 & 0.00 & 12.81 \\
        Median household income (in \$1000s) & 912 & 138.06 & 64.61 & 12.00 & 250.00 \\
        Race: White population (\%) & 912 & 61.16 & 26.09 & 0.00 & 100.00 \\
        Race: Black or African American population (\%) & 912 & 8.55 & 8.55 & 0.00 & 95.10 \\
        Race: Asian population (\%) & 912 & 12.14 & 13.00 & 0.00 & 87.23 \\
        Unemployment rate (\%) & 912 & 5.84 & 6.65 & 0.00 & 53.85 \\
        Walkability index score & 912 & 14.54 & 2.72 & 5.00 & 19.67 \\
        Population density (people/km$^2$) & 769 & 120.45 & 85.54 & 0.00 & 566.00 \\
        Road density & 912 & 3.97 & 6.63 & 0.00 & 71.50 \\
        Intersection density & 912 & 16.13 & 44.47 & 0.00 & 525.44 \\
        Activity density & 912 & 246.66 & 324.42 & 0.00 & 2437.36 \\
        Road Network & 912 & 32.41 & 11.82 & 0.00 & 112.91 \\     
        \bottomrule
    \end{tabular}%
    }
\end{table}

\begin{table}[h!]
    \centering
    \renewcommand{\arraystretch}{1.2}
    \caption{Descriptive statistics for staggered adoption matching pairs in California Bay Area}
    \resizebox{\textwidth}{!}{%
    \begin{tabular}{lccccc}
        \toprule
        \textbf{Variable} & \textbf{N} & \textbf{Mean} & \textbf{Std. dev} & \textbf{Min} & \textbf{Max} \\
        \midrule
        \multicolumn{6}{l}{\textbf{Treated group}} \\
        Monthly number of customers per POI (in 1000s) & 9690 & 0.18 & 0.50 & 0.01 & 20.17 \\
        Median household income (in \$1000s) & 9690 & 102.49 & 58.29 & 0.00 & 250.00 \\
        Race: White population (\%) & 9690 & 45.75 & 22.41 & 0.00 & 97.23 \\
        Race: Black or African American population (\%) & 9690 & 5.64 & 9.61 & 0.00 & 100.00 \\
        Race: Asian population (\%) & 9690 & 21.88 & 19.26 & 0.00 & 95.90 \\
        Unemployment rate (\%) & 9690 & 5.80 & 5.45 & 0.00 & 44.44 \\
        Walkability index score & 9690 & 15.47 & 2.89 & 2.33 & 20.00 \\
        Population density (people/km$^2$) & 9690 & 16.64 & 17.29 & 0.00 & 210.07 \\
        Road density & 9690 & 2.14 & 3.63 & 0.00 & 30.71 \\
        Intersection density (intersections/km$^2$) & 9690 & 5.86 & 11.59 & 0.00 & 120.81 \\
        Activity density & 9690 & 23.99 & 35.70 & 0.09 & 775.45 \\
        Road network & 9690 & 25.96 & 8.50 & 1.48 & 61.35 \\
        \midrule
        \multicolumn{6}{l}{\textbf{Matched control group}} \\
        Monthly number of customers per POI (in 1000s) & 6461 & 0.17 & 0.43 & 0.01 & 15.57 \\
        Median household income (in \$1000s) & 6461 & 102.82 & 58.23 & 0.00 & 250.00 \\
        Race: White population (\%) & 6461 & 45.65 & 21.61 & 0.00 & 97.23 \\
        Race: Black or African American population (\%) & 6461 & 5.60 & 9.65 & 0.00 & 100.00 \\
        Race: Asian population (\%) & 6461 & 22.00 & 19.17 & 0.00 & 95.90 \\
        Unemployment rate (\%) & 6461 & 5.88 & 5.48 & 0.00 & 42.60 \\
        Walkability index score & 6461 & 15.47 & 2.83 & 2.33 & 20.00 \\
        Population density (people/km$^2$) & 6461 & 16.25 & 16.70 & 0.00 & 210.07 \\
        Road density & 6461 & 2.17 & 3.61 & 0.00 & 30.71 \\
        Intersection density & 6461 & 6.14 & 11.99 & 0.00 & 120.81 \\
        Activity density & 6461 & 23.76 & 35.25 & 0.09 & 667.78 \\
        Road Network & 6461 & 25.87 & 8.25 & 0.00 & 61.35 \\     
        \bottomrule
    \end{tabular}%
    }
\end{table}

\begin{table}[h!]
    \centering
    \renewcommand{\arraystretch}{1.2}
    \caption{Descriptive statistics for non-staggered matching pairs in NYC}
    \resizebox{\textwidth}{!}{%
    \begin{tabular}{lccccc}
        \toprule
        \textbf{Variable} & \textbf{N} & \textbf{Mean} & \textbf{Std. Dev} & \textbf{Min} & \textbf{Max} \\
        \midrule
        \multicolumn{6}{l}{\textbf{Treated group}} \\
        Monthly number of customers per POI (in 1000s) & 2066 & 0.46 & 1.81 & 0.00 & 56.57 \\
        Median household income (in \$1000s)  & 2066 & 117.39 & 72.36 & 10.00 & 250.00 \\
        Race: White population (\%) & 2066 & 58.72 & 26.04 & 0.00 & 100.00 \\
        Race: Black or African American population (\%) & 2066 & 10.47 & 18.95 & 0.00 & 100.00 \\
        Race: Asian population (\%) & 2066 & 12.26 & 12.14 & 0.00 & 87.23 \\
        Unemployment Rate (\%) & 2066 & 6.00 & 6.54 & 0.00 & 53.85 \\
        Walkability index score & 2066 & 14.79 & 2.68 & 5.00 & 19.67 \\
        Population density (people/km$^2$) & 2066 & 115.44 & 80.44 & 0.00 & 566.00 \\
        Road density & 2066 & 3.07 & 6.19 & 0.00 & 78.52 \\
        Intersection density & 2066 & 12.91 & 38.39 & 0.00 & 525.44 \\
        Activity density & 2066 & 223.32 & 301.35 & 0.00 & 2437.36 \\
        Road network & 2066 & 32.07 & 11.26 & 0.00 & 112.91 \\
        \midrule
        \multicolumn{6}{l}{\textbf{Matched control group}} \\
        Monthly number of customers per POI (in 1000s) & 1487 & 0.43 & 1.62 & 0.00 & 40.53\\
        Median household income (in \$1000s) & 1487 & 87.70 & 61.48 & 9.02 & 250.00 \\
        Race: White population (\%) & 1487 & 47.85 & 27.27 & 0.00 & 100.00 \\
        Race: Black or African American population (\%) & 1487 & 16.86 & 23.27 & 0.00 & 100.00 \\
        Race: Asian population (\%) & 1487 & 13.89 & 16.40 & 0.00 & 96.11 \\
        Unemployment rate (\%) & 1487 & 7.20 & 7.74 & 0.00 & 76.67 \\
        Walkability index score & 1487 & 15.01 & 2.43 & 7.67 & 19.83 \\
        Population density (people/km$^2$) & 1487 & 99.19 & 74.24 & 0.00 & 430.66 \\
        Road density & 1487 & 2.40 & 6.08 & 0.00 & 42.98 \\
        Intersection density & 1487 & 11.03 & 34.14 & 0.00 & 249.13 \\
        Activity density & 1487 & 161.19 & 270.67 & 0.00 & 2437.36 \\
        Road network & 1487 & 31.45 & 11.02 & 0.00 & 107.34 \\
        \bottomrule
    \end{tabular}
    }
\end{table}

\begin{table}[h!]
    \centering
    \renewcommand{\arraystretch}{1.2}
    \caption{Descriptive statistics for non-staggered matching pairs in California Bay Area}
    \resizebox{\textwidth}{!}{%
    \begin{tabular}{lccccc}
        \toprule
        \textbf{Variable} & \textbf{N} & \textbf{Mean} & \textbf{Std. dev} & \textbf{Min} & \textbf{Max} \\
        \midrule
        \multicolumn{6}{l}{\textbf{Treated group}} \\
        Monthly number of customers per POI (in 1000s) & 10601 & 0.16 & 0.42 & 0.01 & 18.47 \\
        Median household income (in \$1000s) & 10601 & 103.54 & 59.04 & 0.00 & 250.00 \\
        Race: White population (\%) & 10601 & 45.75 & 22.48 & 0.00 & 97.23 \\
        Race: Black or African American population (\%) & 10601 & 5.55 & 9.41 & 0.00 & 89.83 \\
        Race: Asian population (\%) & 10601 & 21.88 & 19.32 & 0.00 & 95.90 \\
        Unemployment rate (\%) & 10601 & 5.83 & 5.48 & 0.00 & 44.44 \\
        Walkability index score & 10601 & 15.47 & 2.89 & 2.33 & 20.00 \\
        Population density (people/km$^2$) & 10601 & 16.73 & 17.78 & 0.00 & 210.07 \\
        Road density & 10601 & 2.11 & 3.56 & 0.00 & 30.71 \\
        Intersection density (intersections/km$^2$) & 10601 & 5.75 & 11.38 & 0.00 & 120.81 \\
        Activity density & 10601 & 24.10 & 35.95 & 0.09 & 775.45 \\
        Road network & 10601 & 25.95 & 8.43 & 1.48 & 61.35 \\
        \midrule
        \multicolumn{6}{l}{\textbf{Matched control group}} \\
        Monthly number of customers per POI (in 1000s) & 8269 & 0.16 & 0.41 & 0.01 & 16.83 \\
        Median household income (in \$1000s) & 8269 & 102.28 & 56.73 & 0.00 & 250.00 \\
        Race: White population (\%) & 8269 & 41.19 & 22.09 & 0.00 & 98.20 \\
        Race: Black or African American population (\%) & 8269 & 6.02 & 11.08 & 0.00 & 97.67 \\
        Race: Asian population (\%) & 8269 & 20.56 & 20.07 & 0.00 & 95.25 \\
        Unemployment rate (\%) & 8269 & 5.78 & 5.22 & 0.00 & 44.55 \\
        Walkability index score & 8269 & 14.45 & 2.82 & 1.67 & 19.83 \\
        Population density (people/km$^2$) & 8269 & 17.94 & 14.01 & 0.00 & 168.74 \\
        Road density & 8269 & 2.07 & 3.93 & 0.00 & 39.53 \\
        Intersection density & 8269 & 5.55 & 11.41 & 0.00 & 146.44 \\
        Activity density & 8269 & 13.24 & 10.45 & 0.20 & 136.29 \\
        Road Network & 8269 & 25.09 & 7.96 & 2.57 & 67.43 \\     
        \bottomrule
    \end{tabular}%
    }
\end{table}

\begin{table}[h!]
    \centering
    \renewcommand{\arraystretch}{1.2}
    \caption{Descriptive statistics for non-staggered PSM pairs in NYC}
    \resizebox{\textwidth}{!}{%
    \begin{tabular}{lccccc}
        \toprule
        \textbf{Variable} & \textbf{N} & \textbf{Mean} & \textbf{Std. Dev} & \textbf{Min} & \textbf{Max} \\
        \midrule
        \multicolumn{6}{l}{\textbf{Treated group}} \\
        Monthly number of customers per POI (in 1000s) & 1872 & 0.44 & 1.43 & 0.00 & 25.90 \\
        Median household income (in \$1000s) & 1872 & 120.59 & 73.45 & 0.00 & 250.00 \\
        Race: White population (\%) & 1872 & 59.60 & 23.81 & 0.00 & 100.00 \\
        Race: Black or African American population (\%) & 1872 & 9.23 & 16.01 & 0.00 & 100.00 \\
        Race: Asian population (\%) & 1872 & 11.80 & 18.58 & 0.00 & 73.80 \\
        Unemployment Rate (\%) & 1872 & 5.88 & 6.23 & 0.00 & 53.85 \\
        Walkability index score & 1872 & 14.73 & 2.68 & 5.00 & 19.67 \\
        Population density (people/km$^2$) & 1872 & 116.37 & 81.27 & 0.00 & 566.00 \\
        Road density & 1872 & 3.08 & 6.11 & 0.00 & 78.52 \\
        Intersection density & 1872 & 13.07 & 38.62 & 0.00 & 525.44 \\
        Activity density & 1872 & 226.22 & 301.71 & 0.00 & 2437.36 \\
        Road network & 1872 & 31.93 & 11.24 & 0.00 & 112.91 \\
        \midrule
        \multicolumn{6}{l}{\textbf{Matched control group}} \\
        Monthly number of customers per POI (in 1000s) & 1082 & 0.38 & 1.96 & 0.00 & 52.38 \\
        Median household income (in \$1000s) & 1082 & 102.74 & 63.89 & 0.00 & 250.00 \\
        Race: White population (\%) & 1082 & 53.60 & 24.02 & 0.00 & 100.00 \\
        Race: Black or African American population (\%) & 1082 & 12.80 & 11.54 & 0.00 & 100.00 \\
        Race: Asian population (\%) & 1082 & 11.50 & 12.91 & 0.00 & 82.81 \\
        Unemployment rate (\%) & 1082 & 6.45 & 6.35 & 0.00 & 40.63 \\
        Walkability index score & 1082 & 14.91 & 2.40 & 7.67 & 19.67 \\
        Population density (people/km$^2$) & 1082 & 105.37 & 75.33 & 0.00 & 580.67 \\
        Road density & 1082 & 2.34 & 6.11 & 0.00 & 71.50 \\
        Intersection density & 1082 & 10.14 & 34.46 & 0.00 & 525.44 \\
        Activity density & 1082 & 171.03 & 270.35 & 0.00 & 2437.36 \\
        Road network & 1082 & 31.08 & 10.80 & 1.23 & 112.91 \\
        \bottomrule
    \end{tabular}
    }
\end{table}

\begin{table}[h!]
    \centering
    \renewcommand{\arraystretch}{1.2}
    \caption{Descriptive statistics for non-staggered PSM pairs in California Bay Area}
    \resizebox{\textwidth}{!}{%
    \begin{tabular}{lccccc}
        \toprule
        \textbf{Variable} & \textbf{N} & \textbf{Mean} & \textbf{Std. dev} & \textbf{Min} & \textbf{Max} \\
        \midrule
        \multicolumn{6}{l}{\textbf{Treated group}} \\
        Monthly number of customers per POI (in 1000s) & 8640 & 0.16 & 0.42 & 0.01 & 18.47 \\
        Median household income (in \$1000s) & 8640 & 103.54 & 59.04 & 0.00 & 250.00 \\
        Race: White population (\%) & 8640 & 45.75 & 22.48 & 0.00 & 97.23 \\
        Race: Black or African American population (\%) & 8640 & 5.55 & 9.41 & 0.00 & 89.83 \\
        Race: Asian population (\%) & 8640 & 21.88 & 19.32 & 0.00 & 95.90 \\
        Unemployment rate (\%) & 8640 & 5.83 & 5.48 & 0.00 & 44.44 \\
        Walkability index score & 8640 & 15.47 & 2.89 & 2.33 & 20.00 \\
        Population density (people/km$^2$) & 8640 & 16.73 & 17.78 & 0.00 & 210.07 \\
        Road density & 8640 & 2.11 & 3.56 & 0.00 & 30.71 \\
        Intersection density (intersections/km$^2$) & 8640 & 5.75 & 11.38 & 0.00 & 120.81 \\
        Activity density & 8640 & 24.10 & 35.95 & 0.09 & 775.45 \\
        Road network & 8640 & 25.95 & 8.43 & 1.48 & 61.35 \\
        \midrule
        \multicolumn{6}{l}{\textbf{Matched control group}} \\
        Monthly number of customers per POI (in 1000s) & 2172 & 0.17 & 0.43 & 0.00 & 14.30 \\
        Median household income (in \$1000s) & 2172 & 108.44 & 53.42 & 0.00 & 250.00 \\
        Race: White population (\%) & 2172 & 44.89 & 21.53 & 0.00 & 97.77 \\
        Race: Black or African American population (\%) & 2172 & 6.38 & 10.30 & 0.00 & 89.24 \\
        Race: Asian population (\%) & 2172 & 20.77 & 10.30 & 0.00 & 92.25 \\
        Unemployment rate (\%) & 2172 & 5.88 & 5.20 & 0.00 & 42.60 \\
        Walkability index score & 2172 & 14.48 & 3.08 & 2.33 & 19.83 \\
        Population density (people/km$^2$) & 2172 & 15.29 & 13.20 & 0.00 & 117.07 \\
        Road density & 2172 & 2.16 & 3.66 & 0.00 & 37.50 \\
        Intersection density & 2172 & 6.23 & 12.08 & 0.00 & 146.44 \\
        Activity density & 2172 & 19.79 & 19.06 & 0.04 & 149.53 \\
        Road Network & 2172 & 24.21 & 8.23 & 1.44 & 59.90 \\     
        \bottomrule
    \end{tabular}%
    }
\end{table}

\begin{table}[h!]
\centering
\caption{Effects of EVCS on customer counts: comparing baseline model results with non-staggered matching, non-staggered PSM, and placebo test results for NYC}
\label{tab:my_table6}
\renewcommand{\arraystretch}{1.1} % Adjust row height
\begin{tabular}{l c c c c}
\hline
  & \textbf{\textit{Baseline}} & \textbf{\textit{Non-staggered}} & \textbf{\textit{Non-staggered PSM}} & \textbf{\textit{Placebo test}}\\
\multirow{2}{*}{}  &  \textbf{\textit{model}} &  \textbf{\textit{matching}} & \textbf{\textit{matching}} & \textbf{\textit{results}} \\
\hline
\multirow{3}{*}{Treatment effect} & $28.52^{\ast\ast\ast}$ & $24.93$ & $-3.62$ & $8.44$ \\
\cline{2-5}
 & (4.59) & (16.21) & (8.63) & (5.32) \\
\cline{2-5}
 & $p=0.000$ & $p=0.124$ & $p=0.675$ & $p=0.112$ \\
\hline
\textbf{Fixed-effects} &   &   &   &   \\
Individual POI & Yes & Yes & Yes & Yes \\
Year & Yes & Yes & Yes & Yes \\
\hline
\textbf{Fit statistics} &   &   &   &   \\
Observations & 6932 & 8364 & 7488 & 7268 \\
R\textsuperscript{2}  & 0.969 & 0.952 & 0.996 & 0.990  \\
\hline
\end{tabular}
\begin{flushleft}
\textit{Note.} Clustered (at the POI level) standard errors are reported in parentheses, and \textit{p}-values from two-sided \textit{t}-tests are listed under standard errors. The dependent variable is the number of customers. *** $p < 0.001$; ** $p < 0.01$; * $p < 0.05$.
\end{flushleft}
\end{table}

\begin{table}[h!]
\centering
\caption{Effects of EVCS on customer counts: comparing baseline model results with non-staggered matching, non-staggered PSM, and placebo test results for California Bay Area}
\label{tab:my_table7}
\renewcommand{\arraystretch}{1.1} % Adjust row height
\begin{tabular}{l c c c c}
\hline
  & \textbf{\textit{Baseline}} & \textbf{\textit{Non-staggered}} & \textbf{\textit{Non-staggered PSM}} & \textbf{\textit{Placebo test}}\\
\multirow{2}{*}{}  &  \textbf{\textit{model}} &  \textbf{\textit{matching}} & \textbf{\textit{matching}} & \textbf{\textit{results}} \\
\hline
\multirow{3}{*}{Treatment effect} & $7.93^{\ast\ast\ast}$ & $7.17^{\ast\ast\ast}$ & $7.94^{\ast\ast\ast}$ & $1.28$ \\
\cline{2-5}
 & (1.25) & (0.80) & (0.315) & (0.70) \\
\cline{2-5}
 & $p=0.000$ & $p=0.000$ & $p=0.000$ & $p=0.066$ \\
\hline
\textbf{Fixed-effects} &   &   &   &   \\
Individual POI & Yes & Yes & Yes & Yes \\
Year & Yes & Yes & Yes & Yes \\
\hline
\textbf{Fit statistics} &   &   &   &   \\
Observations & 38760 & 42404 & 34560 & 36192 \\
R\textsuperscript{2}  & 0.985 & 0.991 & 0.997 & 0.994  \\
\hline
\end{tabular}
\begin{flushleft}
\textit{Note.} Clustered (at the POI level) standard errors are reported in parentheses, and \textit{p}-values from two-sided \textit{t}-tests are listed under standard errors. The dependent variable is the number of customers. *** $p < 0.001$; ** $p < 0.01$; * $p < 0.05$.
\end{flushleft}
\end{table}

\begin{table}[h!]
\centering
\caption{Heterogeneity by POI categories: comparing baseline model results with non-staggered matching, non-staggered PSM, and placebo test results for NYC}
\label{tab:my_table8}
\renewcommand{\arraystretch}{1.0} % Adjust row height
\resizebox{\textwidth}{!}{
\begin{tabular}{p{6.2cm} c c c c} % Adjust width of the first column
\hline
  & \textbf{\textit{Baseline}} & \textbf{\textit{Non-staggered}} & \textbf{\textit{Non-staggered PSM}} & \textbf{\textit{Placebo test}} \\
\multirow{2}{*}{}  &  \textbf{\textit{model}} &  \textbf{\textit{matching}} &  \textbf{\textit{matching}} & \textbf{\textit{results}} \\
\hline
\multicolumn{5}{l}{\textbf{Treatment effects in:}} \\
\textbf{Apparel} & $28.45$ & $27.04$ & $14.15$ & $-6.04$ \\
 & (24.07) & (81.88) & (41.12) & (29.85) \\
 & $p=0.237$ & $p=0.741$ & $p=0.731$ & $p=0.840$ \\
\textbf{Beauty and Spa} & $9.67$ & $-1.29$ & $0.19$ & $5.16$ \\
 & (22.66) & (85.01) & (40.37) & (28.15) \\
 & $p=0.670$ & $p=0.988$ & $p=0.996$ & $p=0.855$ \\
\textbf{Buildings and Ent. Centers} & $85.73^{\ast\ast\ast}$ & $80.84$ & $-10.21$ & $-23.65$ \\
 & (17.98) & (65.43) & (36.60) & (20.65) \\
 & $p=0.000$ & $p=0.217$ & $p=0.780$ & $p=0.252$ \\
\textbf{Dining} & $14.94^{\ast}$ & $9.27$ & $24.70$ & $7.86$ \\
 & (7.33) & (27.90) & (14.42) & (8.59) \\
 & $p=0.041$ & $p=0.740$ & $p=0.087$ & $p=0.360$ \\
\textbf{Groceries} & $2.57$ & $2.29$ & $6.19$ & $1.41$ \\
 & (20.01) & (65.94) & (34.15) & (23.71) \\
 & $p=0.898$ & $p=0.972$ & $p=0.856$ & $p=0.953$ \\
\textbf{Home Impro. and Furnishings} & $2.81$ & $-5.10$ & $-4.23$ & $1.59$ \\
 & (31.01) & (102.47) & (56.09) & (35.04) \\
 & $p=0.928$ & $p=0.960$ & $p=0.940$ & $p=0.964$ \\
\textbf{Hotels and Casinos} & $45.81^{\ast}$ & $28.79$ & $-46.06$ & $51.79^{\ast\ast}$ \\
 & (20.11) & (70.18) & (38.14) & (21.99) \\
 & $p=0.023$ & $p=0.682$ & $p=0.227$ & $p=0.019$ \\
\textbf{Leisure} & $67.76^{\ast\ast\ast}$ & $52.62$ & $37.66$ & $-20.82$ \\
 & (11.44) & (43.14) & (22.26) & (13.96) \\
 & $p=0.000$ & $p=0.223$ & $p=0.091$ & $p=0.136$ \\
\textbf{Medical and Health} & $1.18$ & $-4.03$ & $28.07$ & $42.95^{\ast}$ \\
 & (24.33) & (75.74) & (35.90) & (24.88) \\
 & $p=0.961$ & $p=0.958$ & $p=0.434$ & $p=0.084$ \\
\textbf{Shopping Centers} & $117.83^{\ast\ast\ast}$ & $279.83^{\ast\ast}$ & $74.55$ & $179.11^{\ast\ast\ast}$ \\
 & (35.11) & (101.52) & (56.09) & (33.24) \\
 & $p=0.001$ & $p=0.006$ & $p=0.843$ & $p=0.000$ \\
\textbf{Shops and Services} & $6.39$ & $-1.67$ & $-12.17$ & $3.54$ \\
 & (11.52) & (41.90) & (21.83) & (13.23) \\
 & $p=0.579$ & $p=0.968$ & $p=0.577$ & $p=0.789$ \\
\hline
\textbf{Fixed-effects} &   &   &   &   \\
Individual POI & Yes & Yes & Yes & Yes \\
Year & Yes & Yes & Yes & Yes \\
\hline
\textbf{Fit statistics} &   &   &   &   \\
Observations & 6932 & 8364 & 7488 & 7268 \\
R\textsuperscript{2}  & 0.971 & 0.954 & 0.997 & 0.991  \\
\hline
\end{tabular}
}
\begin{flushleft}
\textit{Note.} Clustered (at the POI level) standard errors are reported in parentheses, and \textit{p}-values from two-sided \textit{t}-tests are listed under standard errors. The dependent variable is the number of customers. *** $p < 0.001$; ** $p < 0.01$; * $p < 0.05$.
\end{flushleft}
\end{table}

\begin{table}[h!]
\centering
\caption{Heterogeneity by POI categories: comparing baseline model results with non-staggered matching, non-staggered PSM, and placebo test results for California Bay Area}
\label{tab:my_table9}
\renewcommand{\arraystretch}{1.0}
\resizebox{\textwidth}{!}{
\begin{tabular}{p{6.2cm} c c c c}
\hline
 & \textbf{\textit{Baseline}} & \textbf{\textit{Non-staggered}} & \textbf{\textit{Non-staggered PSM}} & \textbf{\textit{Placebo test}} \\
\multirow{2}{*}{} & \textbf{\textit{model}} & \textbf{\textit{matching}} & \textbf{\textit{matching}} & \textbf{\textit{results}} \\
\hline
\multicolumn{5}{l}{\textbf{Treatment effects in:}} \\

\textbf{Shopping Centers} & $21.02$ & $18.25$ & $13.95$ & $-2.05$ \\
 & (11.56) & (9.52) & (7.44) & (2.96) \\
 & $p=0.237$ & $p=0.095$ & $p=0.081$ & $p=0.840$ \\
\textbf{Buildings and Ent. Centers} & $-1.93$ & $-0.76$ & $3.47$ & $1.80$ \\
 & (4.05) & (2.59) & (2.95) & (6.63) \\
 & $p=0.634$ & $p=0.769$ & $p=0.240$ & $p=0.786$ \\
\textbf{Leisure} & $1.03$ & $5.59$ & $6.63^{\ast\ast\ast}$ & $5.21$ \\
 & (3.54) & (4.23) & (1.27) & (3.09) \\
 & $p=0.770$ & $p=0.187$ & $p=0.000$ & $p=0.092$ \\
\textbf{Hotels and Casinos} & $2.03$ & $14.90$ & $23.20^{\ast\ast\ast}$ & $3.26$ \\
 & (10.30) & (7.86) & (3.50) & (5.03) \\
 & $p=0.770$ & $p=0.058$ & $p=0.000$ & $p=0.517$ \\
\textbf{Apparel} & $7.31^{\ast}$ & $12.21^{\ast\ast\ast}$ & $6.46^{\ast\ast\ast}$ & $1.85$ \\
 & (3.15) & (3.39) & (1.62) & (2.49) \\
 & $p=0.021$ & $p=0.000$ & $p=0.000$ & $p=0.455$ \\
\textbf{Dining} & $7.70^{\ast\ast\ast}$ & $8.26^{\ast\ast\ast}$ & $6.96^{\ast\ast\ast}$ & $7.04^{\ast\ast\ast}$ \\
 & (1.03) & (1.23) & (0.74) & (1.19) \\
 & $p=0.000$ & $p=0.000$ & $p=0.000$ & $p=0.000$ \\
\textbf{Beauty and Spa} & $1.76^{\ast}$ & $3.79^{\ast\ast\ast}$ & $2.44^{\ast\ast\ast}$ & $0.88$ \\
 & (0.87) & (0.85) & (0.51) & (2.17) \\
 & $p=0.044$ & $p=0.000$ & $p=0.000$ & $p=0.685$ \\
\textbf{Shops and Services} & $4.89^{\ast\ast\ast}$ & $3.17^{\ast\ast}$ & $4.55^{\ast\ast\ast}$ & $1.73$ \\
 & (0.80) & (0.92) & (0.57) & (1.22) \\
 & $p=0.000$ & $p=0.001$ & $p=0.000$ & $p=0.157$ \\
\textbf{Home Impro. and Furnishings} & $0.94$ & $1.15$ & $6.10^{\ast\ast}$ & $-1.52$ \\
 & (1.36) & (2.25) & (2.05) & (2.98) \\
 & $p=0.490$ & $p=0.610$ & $p=0.003$ & $p=0.609$ \\
\textbf{Groceries} & $9.93^{\ast}$ & $2.79$ & $11.58^{\ast\ast\ast}$ & $1.10$ \\
 & (4.27) & (6.54) & (3.10) & (3.67) \\
 & $p=0.020$ & $p=0.669$ & $p=0.000$ & $p=0.763$ \\
\textbf{Medical and Health} & $4.32^{\ast}$ & $3.51^{\ast}$ & $5.46^{\ast\ast\ast}$ & $1.74$ \\
 & (1.93) & (1.40) & (1.45) & (2.45) \\
 & $p=0.025$ & $p=0.012$ & $p=0.000$ & $p=0.476$ \\
\hline
\textbf{Fixed-effects} &   &   &   &   \\
Individual POI & Yes & Yes & Yes & Yes \\
Year & Yes & Yes & Yes & Yes \\
\hline
\textbf{Fit statistics} &   &   &   &   \\
Observations & 36072 & 36776 & 34560 & 32192 \\
R\textsuperscript{2}  & 0.974 & 0.992 & 0.998 & 0.994 \\
\hline
\end{tabular}
}
\begin{flushleft}
\textit{Note.} Clustered (at the POI level) standard errors are reported in parentheses, and \textit{p}-values from two-sided \textit{t}-tests are listed under standard errors. The dependent variable is the number of customers. *** $p < 0.001$; ** $p < 0.01$; * $p < 0.05$.
\end{flushleft}
\end{table}

\begin{table}[h!]
\centering
\caption{Variation in treatment effects by distance: comparing baseline model results with non-staggered matching, non-staggered PSM, and placebo test results for NYC}
\label{tab:my_table10}
\renewcommand{\arraystretch}{1.1} % Adjust row height
\begin{tabular}{p{3.5cm} c c c c} % Adjust width of the first column
\hline
  & \textbf{\textit{Baseline}} & \textbf{\textit{Non-staggered}} & \textbf{\textit{Non-staggered PSM}} & \textbf{\textit{Placebo test}} \\
\multirow{2}{*}{}  &  \textbf{\textit{model}} &  \textbf{\textit{matching}} & \textbf{\textit{matching}} & \textbf{\textit{results}} \\
\hline
\multicolumn{5}{l}{\textbf{Treatment effects in:}} \\
\textbf{0-100m} & $39.12^{\ast}$ & $34.88$ & $-40.014$ & $8.71$ \\
 & (16.61) & (55.60) & (28.69) & (11.31) \\
 & $p=0.019$ & $p=0.530$ & $p=0.163$ & $p=0.622$ \\
\textbf{100-200m} & $25.14^{\ast}$ & $-6.12$ & $-63.57^{\ast\ast\ast}$ & $9.84$ \\
 & (9.93) & (36.31) & (19.23) & (11.31) \\
 & $p=0.011$ & $p=0.866$ & $p=0.001$ & $p=0.385$ \\
\textbf{200-300m} & $27.04^{\ast\ast\ast}$ & $21.36$ & $11.31$ & $16.67$ \\
 & (7.21) & (26.06) & (13.81) & (8.53) \\
 & $p=0.000$ & $p=0.413$ & $p=0.413$ & $p=0.051$ \\
\textbf{300-400m} & $30.10^{\ast\ast\ast}$ & $48.52$ & $26.13$ & $-2.64$ \\
 & (8.01) & (29.37) & (15.25) & (9.41) \\
 & $p=0.000$ & $p=0.099$ & $p=0.087$ & $p=0.779$ \\
\hline
\textbf{Fixed-effects} &   &   &   &   \\
Individual POI & Yes & Yes & Yes & Yes \\
Year & Yes & Yes & Yes & Yes \\
\hline
\textbf{Fit statistics} &   &   &   &   \\
Observations & 6932 & 8364 & 7488 & 7268 \\
R\textsuperscript{2}  & 0.969 & 0.952 & 0.997 & 0.990  \\
\hline
\end{tabular}
\begin{flushleft}
\textit{Note.} Clustered (at the POI level) standard errors are reported in parentheses, and \textit{p}-values from two-sided \textit{t}-tests are listed under standard errors. The dependent variable is the number of customers. *** $p < 0.001$; ** $p < 0.01$; * $p < 0.05$.
\end{flushleft}
\end{table}

\begin{table}[h!]
\centering
\caption{Variation in treatment effects by distance: comparing baseline model results with non-staggered matching, non-staggered PSM, and placebo test results for California Bay Area}
\label{tab:my_table11}
\renewcommand{\arraystretch}{1.1} % Adjust row height
\begin{tabular}{p{3.5cm} c c c c} % Adjust width of the first column
\hline
  & \textbf{\textit{Baseline}} & \textbf{\textit{Non-staggered}} & \textbf{\textit{Non-staggered PSM}} & \textbf{\textit{Placebo test}} \\
\multirow{2}{*}{}  &  \textbf{\textit{model}} &  \textbf{\textit{matching}} & \textbf{\textit{matching}} & \textbf{\textit{results}} \\
\hline
\multicolumn{5}{l}{\textbf{Treatment effects in:}} \\
\textbf{0-100m} & $5.58^{\ast}$ & $5.93^{\ast\ast\ast}$ & $7.07^{\ast\ast\ast}$ & $2.07$ \\
 & (2.35) & (1.65) & (1.19) & (7.80) \\
 & $p=0.017$ & $p=0.000$ & $p=0.000$ & $p=0.791$ \\
\textbf{100-200m} & $9.74^{\ast}$ & $6.50^{\ast\ast\ast}$ & $5.28^{\ast\ast\ast}$ & $1.62$ \\
 & (4.10) & (1.24) & (1.28) & (1.28) \\
 & $p=0.018$ & $p=0.000$ & $p=0.000$ & $p=0.207$ \\
\textbf{200-300m} & $5.49^{\ast\ast\ast}$ & $6.52^{\ast\ast}$ & $7.46^{\ast\ast\ast}$ & $2.77^{\ast\ast}$ \\
 & (1.26) & (2.10) & (0.82) & (0.93) \\
 & $p=0.000$ & $p=0.002$ & $p=0.000$ & $p=0.003$ \\
\textbf{300-400m} & $4.62^{\ast\ast}$ & $9.56^{\ast\ast\ast}$ & $5.99^{\ast\ast\ast}$ & $-0.46$ \\
 & (1.45) & (2.07) & (0.77) & (0.93) \\
 & $p=0.001$ & $p=0.000$ & $p=0.000$ & $p=0.619$ \\
\hline
\textbf{Fixed-effects} &   &   &   &   \\
Individual POI & Yes & Yes & Yes & Yes \\
Year & Yes & Yes & Yes & Yes \\
\hline
\textbf{Fit statistics} &   &   &   &   \\
Observations & 35356 & 44004 & 35812 & 22036 \\
R\textsuperscript{2}  & 0.982 & 0.991 & 0.998 & 0.992 \\
\hline
\end{tabular}
\begin{flushleft}
\textit{Note.} Clustered (at the POI level) standard errors are reported in parentheses, and \textit{p}-values from two-sided \textit{t}-tests are listed under standard errors. The dependent variable is the number of customers. *** $p < 0.001$; ** $p < 0.01$; * $p < 0.05$.
\end{flushleft}
\end{table}

\begin{table}[h!]
\centering
\caption{Variation in treatment effects across different income groups: comparing baseline model results with non-staggered matching, non-staggered PSM, and placebo test results for NYC}
\label{tab:my_table12}
\renewcommand{\arraystretch}{1.0} % Adjust row height
\resizebox{\textwidth}{!}{
\begin{tabular}{p{4.5cm} c c c c} % Adjust width of the first column
\hline
  & \textbf{\textit{Baseline}} & \textbf{\textit{Non-staggered}} & \textbf{\textit{Non-staggered PSM}} & \textbf{\textit{Placebo}} \\
\multirow{2}{*}{}  &  \textbf{\textit{model}} &  \textbf{\textit{matching approach}} &  \textbf{\textit{approach}} & \textbf{\textit{test results}} \\
\hline
\multicolumn{5}{l}{\textbf{Treatment effects in:}} \\
\textbf{$<\$60k$} & $23.76^{\ast}$ & $20.05$ & $0.59$ & $9.29$ \\
 & (11.79) & (14.27) & (16.44) & (13.24) \\
 & $p=0.044$ & $p=0.160$ & $p=0.972$ & $p=0.483$ \\
\textbf{\$60K-\$100K} & $34.35^{\ast\ast\ast}$ & $23.55$ & $27.43^{\ast}$ & $18.75$ \\
 & (9.23) & (12.54) & (13.14) & (10.23) \\
 & $p=0.000$ & $p=0.064$ & $p=0.037$ & $p=0.064$ \\
\textbf{\$100K-\$150K} & $42.43^{\ast\ast}$ & $35.62$ & $22.49$ & $19.64$ \\
 & (16.16) & (18.86) & (23.47) & (18.05) \\
 & $p=0.009$ & $p=0.099$ & $p=0.338$ & $p=0.277$ \\
\textbf{$>\$150k$} & $37.32^{\ast\ast\ast}$ & $37.20^{\ast\ast\ast}$ & $24.64$ & $0.69$ \\
 & (9.06) & (11.51) & (13.24) & (9.89) \\
 & $p=0.000$ & $p=0.001$ & $p=0.063$ & $p=0.944$ \\
\hline
\textbf{Fixed-effects} &   &   &   & \\
Individual POI & Yes & Yes & Yes & Yes \\
Year & Yes & Yes & Yes & Yes \\
\hline
\textbf{Fit statistics} &   &   &   & \\
Observations & 6932 & 8364 & 7488 & 7268 \\
R\textsuperscript{2}  & 0.964 & 0.955 & 0.998 & 0.989  \\
\hline
\end{tabular}
}
\begin{flushleft}
\textit{Note.} Clustered (at the POI level) standard errors are reported in parentheses, and \textit{p}-values from two-sided \textit{t}-tests are listed under standard errors. The dependent variable is the number of customers. *** $p < 0.001$; ** $p < 0.01$; * $p < 0.05$.
\end{flushleft}
\end{table}

\begin{table}[h!]
\centering
\caption{Variation in treatment effects across different income groups: comparing baseline model results with non-staggered matching, non-staggered PSM, and placebo test results for California Bay Area}
\label{tab:my_table13}
\renewcommand{\arraystretch}{1.0} % Adjust row height
\resizebox{\textwidth}{!}{
\begin{tabular}{p{4.5cm} c c c c} % Adjust width of the first column
\hline
  & \textbf{\textit{Baseline}} & \textbf{\textit{Non-staggered}} & \textbf{\textit{Non-staggered PSM}} & \textbf{\textit{Placebo}} \\
\multirow{2}{*}{}  &  \textbf{\textit{model}} &  \textbf{\textit{matching approach}} &  \textbf{\textit{approach}} & \textbf{\textit{test results}} \\
\hline
\multicolumn{5}{l}{\textbf{Treatment effects in:}} \\
\textbf{$<\$60k$} & $4.21^{\ast}$ & $4.56^{\ast\ast}$ & $7.05^{\ast\ast\ast}$ & $-0.04$ \\
 & (1.90) & (1.62) & (0.95) & (1.32) \\
 & $p=0.027$ & $p=0.005$ & $p=0.000$ & $p=0.971$ \\
\textbf{\$60K-\$100K} & $3.75^{\ast\ast}$ & $6.50^{\ast\ast\ast}$ & $4.01^{\ast\ast\ast}$ & $2.51^{\ast\ast}$ \\
 & (1.35) & (1.12) & (0.70) & (0.93) \\
 & $p=0.006$ & $p=0.000$ & $p=0.000$ & $p=0.007$ \\
\textbf{\$100K-\$150K} & $7.27^{\ast\ast\ast}$ & $6.16^{\ast\ast\ast}$ & $5.77^{\ast\ast\ast}$ & $1.14$ \\
 & (1.43) & (1.20) & (0.73) & (0.98) \\
 & $p=0.000$ & $p=0.000$ & $p=0.000$ & $p=0.243$ \\
\textbf{$>\$150k$} & $10.02^{\ast\ast\ast}$ & $11.71^{\ast\ast\ast}$ & $9.78^{\ast\ast\ast}$ & $1.74$ \\
 & (1.61) & (1.30) & (0.82 & (1.10) \\
 & $p=0.000$ & $p=0.000$ & $p=0.000$ & $p=0.112$ \\
\hline
\textbf{Fixed-effects} &   &   &   & \\
Individual POI & Yes & Yes & Yes & Yes \\
Year & Yes & Yes & Yes & Yes \\
\hline
\textbf{Fit statistics} &   &   &   & \\
Observations & 35648 & 36776 & 34560 & 19968 \\
R\textsuperscript{2}  & 0.993 & 0.955 & 0.998 & 0.991  \\
\hline
\end{tabular}
}
\begin{flushleft}
\textit{Note.} Clustered (at the POI level) standard errors are reported in parentheses, and \textit{p}-values from two-sided \textit{t}-tests are listed under standard errors. The dependent variable is the number of customers. *** $p < 0.001$; ** $p < 0.01$; * $p < 0.05$.
\end{flushleft}
\end{table}

\clearpage

\bibliographystyle{plain}
\bibliography{sample2}